\documentclass[11pt]{article}
\usepackage{amsfonts,amsmath,amssymb,latexsym,graphicx,hyperref,mciteplus}
\usepackage[english]{babel}

\begin{document}
\input epsf

\makeatletter
\@addtoreset{equation}{section}
\makeatother


\begin{flushright}
BRX-TH-6286
\end{flushright}
\vspace{20mm}

 \begin{center}
{\bf\Large Analyzing the squeezed state generated by a twist deformation}
\\
\vspace{18mm}
{ Benjamin A. Burrington$^a$\footnote{benjamin.a.burrington@hofstra.edu}, Samir D. Mathur$^b$\footnote{mathur@mps.ohio-state.edu}, Amanda W. Peet$^c$\footnote{awpeet@physics.utoronto.ca}, Ida G. Zadeh$^d$\footnote{zadeh@brandeis.edu}
\\}
\vspace{8mm}
\it{$^a$ Department of Physics and Astronomy, Hofstra University, Hempstead, NY 11549, USA\\
\vspace{3mm}
$^b$ Department of Physics, The Ohio State University, Columbus, OH 43210, USA\\
\vspace{3mm}
$^c$ Department of Physics, University of Toronto, Toronto, Ontario M5S 1A7, Canada\\
\vspace{3mm}
$^d$ Martin Fisher School of Physics, Brandeis University, Waltham, Massachusetts, 02454, USA\\
\vspace{8mm}
}
\end{center}
\vspace{10mm}

\thispagestyle{empty}
\begin{abstract}
The D1D5 CFT has provided a useful microscopic model for studying black holes. The coupling in this theory is a twist deformation whose action on the vacuum generates a squeezed state. We give a new derivation of the expression for this squeezed state using the conformal Ward identity; this derivation provides an insight into several features of the state. We also examine the squeezed state in a continuum limit where we describe it in terms of position space correlations created by the twist.
\end{abstract}
\newpage
\renewcommand{\theequation}{\arabic{section}.\arabic{equation}}
\def\nn{\nonumber \\}
\def\p{\partial}
\def\h{{\frac{1}{2}}}
\def\be{\begin{equation}}
\def\bea{\begin{eqnarray}}
\def\ee{\end{equation}}
\def\eea{\end{eqnarray}}
\def\r{\rightarrow}
\def\tildr{\tilde}
\def\n{\nonumber}
\def\nn{\nonumber \\}
\def\t{\tilde}
\def\b{\bigskip}
\def\sone{\sqrt{w_1-w_0}}
\def\stwo{\sqrt{w_2-w_0}}
\newcommand{\Nsc}{\mathcal{N}}
\newcommand{\bj}{\bar{\jmath}}
\def\sqi{{\frac{1}{\sqrt{2}}}}
\renewcommand\eqref[1]{(\ref{#1})}

\newcommand\ket[1]{|#1\rangle}
\newcommand\bra[1]{\langle #1|}
\newcommand\com[2]{[#1,\,#2]}
\newcommand\ac[2]{\{#1,\,#2\}}
\section{Introduction}\label{intro}
\label{intr}\setcounter{equation}{0}

A very useful microscopic model for the study of black holes
\cite{counting,radiation,fuzzballs,adscft,microrad}
has been the D1D5 conformal field theory (CFT) \cite{dmw02}. One makes a bound state of D1 and D5 branes, and the CFT emerges as a description of the low energy dynamics of this bound state.  While this CFT is complicated in general, it has been conjectured that there is a point in its moduli space of couplings where it becomes particularly simple. At this orbifold point the theory is described by a sigma model whose target space is a symmetric orbifold \cite{orbifold}.

At the orbifold point, the CFT is free; using this free theory enables us to extend the Strominger-Vafa computation \cite{stromingervafa95} from extremal to near-extremal black holes \cite{callanmaldacena96}. Such couplings do not describe the point in moduli space where the dual gravity theory is weakly coupled, giving the supergravity approximation.  In order to work towards describing interesting gravity processes such as black hole formation, one therefore needs to turn on an exactly marginal deformation operator in the CFT, which deforms the theory by `blowing up' the orbifold singularities of the target space. The deformation operator is built out of a twist operator of the orbifold theory dressed with a supercharge \cite{dmw02}.

It is interesting and important to study the effect of this deformation operator ${\cal{O}}_D$ on states of the CFT \cite{dmw99pp,lm,acm10a,acm10b,deform1,deform2,deform3,deform4,deform5,deform6}.
\cite{acm10a,acm10b} described the effect of ${\cal{O}}_D$ on the Ramond vacuum, and on states containing one or two initial quanta.
\cite{deform1} considered transitions between different Ramond vacua via absorption and emission of chiral primaries; processes involving the change of angular momentum by $k$ units were found to be suppressed as $1/N^k$.
\cite{deform2,deform3} took a different tack, studying the computation of anomalous dimensions at first order in conformal perturbation theory for low-lying string states in the CFT and operator mixing.
\cite{deform4} studied the effect of ${\cal{O}}_D$ for bosonic fields, when the twist links together winding numbers $M$ and $N$ to winding number $M+N$. \cite{deform5} looked at the effect of ${\cal{O}}_D$ in the limit when excitation wavelengths are short compared to the gap. The method of \cite{deform6} also allows handling fermions.

Overall, it was found that the twist involved in the deformation operator converts the vacuum into a squeezed state, with a schematic form $e^{\gamma^B_{mn}\,\alpha_{-m}\,\alpha_{-n}\,+\,\gamma^F_{mn}\,d_{-m}\,d_{-n}}|0\rangle$.
The coefficients $\gamma^B, \gamma^F$ in the above squeezed state are given by closed form expressions, but the derivation of these expressions was somewhat lengthy. In the present paper we find a much more direct way of obtaining $\gamma^B, \gamma^F$.  In our new method, we consider the OPE of the stress tensor $T$ with the deformation operator ${\cal O}(w)$.  The conformal Ward identity relates this OPE to a derivative $\p_w {\cal O}(w)$. We then find that this derivative has a simple expression, and performing an integral then gives the $\gamma^B_{mn}, \gamma^F_{mn}$.  This derivation also gives insight into the structure of these coefficients: they are seen to be given by an integral of an expression that is a product of factors, one depending only on $m$ and the other depending only on $n$.

One may consider a continuum limit where $m,n\gg 1$; this corresponds to looking at wavelengths much shorter than the `box size' set by the circle on which the D1D5 CFT lives (see figure \ref{contlimitbb}). We obtain the expression of the squeezed state in this limit, by recasting the $\gamma^B, \gamma^F$ as position space kernels; thus the squeezed state is written in terms of 2-point correlations of fields generated after the twist.  The product structure in the derivatives of $\gamma^B$ and $\gamma^F$ mentioned earlier becomes a product structure for the position space kernel found in the continuum limit: the kernel ${\mathcal K}(w_1,w_2;w_0)$ becomes a product of two terms, one depending only on $w_1$ and one depending only on $w_2$.

We organize the paper as follows.  First we summarize earlier work in section 2.  In section 3 we consider the action of the stress tensor to re-derive the form of the squeezed state using the conformal Ward identity.  Finally in section 4, we compute the continuum limit of the squeezed state, expressing it in terms of a position space kernel.  In section 5 we comment on the utility of the stress-tensor method and on the form of the squeezed state.

\section{Summary of earlier results}\label{sectiontwo}

In this section we review some facts about the D1D5 CFT and summarize some results from \cite{acm10a,acm10b}.

\subsection{The D1D5 CFT}\label{ii}

We compactify IIB string theory as $M_{9,1}\rightarrow M_{4,1}\times S^1\times T^4$. Wrapping $N_1$ D1 branes on $S^1$ and $N_5$ D5 branes on $S^1\times T^4$, we get a bound state that is described at low energy by the D1D5 CFT.

It is believed that there is an orbifold point in the moduli space of couplings where the CFT is given by a sigma model with target space the symmetric product of $N_1N_5$ copies of $T^4$ \cite{orbifold}. At this point the CFT is given by $N_1N_5$ copies of a $c=6$ CFT containing 4 free bosons and 4 free fermions.  The fact that we have a symmetric product in the target space implies that we get twist sectors where different copies of the $c=6$ CFT get twisted together to make a $c=6$ CFT living on a longer circle.

The CFT has ${\cal N}=(4,4)$ supersymmetry; we list the algebra generators and their commutators in Appendix \ref{appa}. The ${\cal  N}=4$ superalgebra contains a $SU(2)$ current algebra under which the fermions form doublets, labelled by an index $\alpha$. (This doublet structure is described in eqs. (\ref{aaone}),(\ref{aatwo}).) The four bosons are grouped into representations of the $SO(4)\approx SU_1(2)\times SU_2(2)$ symmetry group of the $T^4$; the doublets under these two $SU(2)$ groups are labelled by indices $A, \dot A$ (eq. (\ref{aathree})).

We consider the Euclidean theory for which the base space is a cylinder with coordinates $\tau$ and $\sigma$ where $-\infty<\tau<\infty$ and $0\le\sigma<2\pi$. We deform the CFT off the orbifold point by the operators
\be\label{fullstate}
\hat O_{\dot A\dot B}(w_0)=\Big [{\frac{1}{2\pi i}} \int _{w_0} dw G^-_{\dot A} (w)\Big ]\Big [{\frac{1}{2\pi i}} \int _{\bar w_0} d\bar w \bar G^-_{\dot B} (\bar w)\Big ]\sigma_2^{++}(w_0).
\ee
The operator $\sigma_2^+(w_0)$ is a twist which links together two copies of the CFT at the point $w_0$; the $+$ superscript indicates that it carries a charge $\h$ under the $SU(2)$ contained in the superalgebra.  $G^-_{\dot A}$ is the supercurrent.

\begin{figure}[ht]
\begin{center}
\includegraphics[width=7cm]{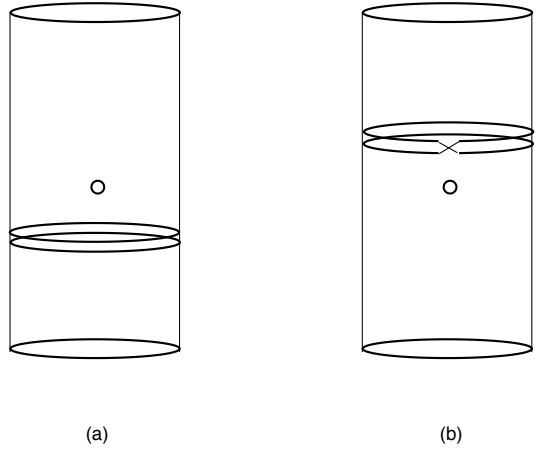}
\caption{The twist operator joins two copies of the CFT on two singly-wound circles into one copy of the CFT on a doubly-wound circle.}\label{2-twist}
\end{center}
\end{figure}

The left and right moving parts of all operators and states are decoupled in our computations and thus in what follows we will write only the left movers. The D1D5 bound state as constructed above gives fermions that are periodic around the $S^1$, so they are in the Ramond sector. We can map Ramond sector states to NS sector states by spectral flow, which is a symmetry that changes dimensions and charges as follows \cite{spectral}
\be
h'=h+\alpha j +{\frac{c\alpha^2}{24}}, ~~~
j'=j+{\frac{c\alpha}{12}}
\label{spectral}
\ee
while the operators themselves change as
\be
T'=T-\alpha J+{\frac{c\alpha^2}{24}}, ~~~J'^3=J^3-{\frac{c\alpha}{12}}
\ee
(Thus we get $\langle\psi|L_0|\psi\rangle=\langle\psi'|L'_0|\psi'\rangle$ and $\langle\psi|J_0|\psi\rangle=\langle\psi'|J'^3_0|\psi'\rangle$). The transformations of the stress-energy tensor and the $R$-current under the spectral flow have been explicitly derived in Appendix \ref{appb}.

In \cite{acm10a,acm10b} the situation in figure (\ref{2-twist}) was studied. We start with two copies of the $c=6$ CFT, each on a singly wound circle. The twist $\sigma_2^+$ joins these copies into one copy of the $c=6$ CFT living on a doubly wound circle. The twist operator is inserted at the point $w_0=\tau_0+i\sigma_0$ on the cylinder.

The Ramond sector ground states are doublets under $SU(2)$, and we choose the initial state
\be
|0_R^{-}\rangle^{(1)}\otimes |0_R^{-}\rangle^{(2)}
\ee
which has spin $-\h$ for each copy. The twist $\sigma_2^+$ is normalized so that
\be
\sigma_2^+(w_0)\;|0_R^-\rangle^{(1)}\otimes |0_R^-\rangle^{(2)}=|0_R^-\rangle+\dots
\ee
where $|0_R^-\rangle$ is the (normalised) negative charge Ramond ground state of the CFT on the doubly wound circle.

\subsection{The mode expansions and the exponential ansatz}\label{ii-i}

We first expand the modes of bosonic and fermionic fields on the cylinder.  For $\tau<\tau_0$, \emph{i.e.} below the twist insertion, these modes are given by
\bea
\alpha^{(i)}_{A\dot A, n}&=& {\frac{1}{2\pi}} \int_{\sigma=0}^{2\pi} \p_w X^{(i)}_{A\dot A}(w) e^{nw} dw, ~~~i=1,2
\label{lone}\\
d^{(i)\alpha A}_n&=&{\frac{1}{ 2\pi i}} \int_{\sigma=0}^{2\pi} \psi^{(i)\alpha A}(w) e^{nw} dw, ~~~i=1,2
\label{loneqq}
\eea
and their (anti-)commutation relations are
\bea
[\alpha^{(i)}_{A\dot A, m}, \alpha^{(j)}_{B\dot B, n}]&=&-\epsilon_{AB}\epsilon_{\dot A\dot B}\delta^{ij} m \delta_{m+n,0},\\
\{ d^{(i)\alpha A}_m, d^{(j)\beta B}_n\}&=&-\epsilon^{\alpha\beta}\epsilon^{AB}\delta^{ij}\delta_{m+n,0}.
\eea
Above the twist insertion ($\tau>\tau_0$)  the CFT lives on a doubly twisted circle and the boson and fermion modes are
\bea
\alpha_{A\dot A, n}&=& {\frac{1}{2\pi}} \int_{\sigma=0}^{4\pi} \p_w X_{A\dot A}(w) e^{{\frac{n}{2}}w} dw,
\label{qaone}\\
d^{\alpha A}_n&=& {\frac{1}{2\pi i}}\int_{\sigma=0}^{4\pi} \psi^{\alpha A}(w) e^{\frac{n}{2} w} dw.
\label{qaoneqq}
\eea
The (anti-)commutation relations for these modes read
\bea\label{bcommtwist}
[\alpha_{A\dot A, m}, \alpha_{B\dot B, n}]&=&-\epsilon_{AB}\epsilon_{\dot A\dot B} m \delta_{m+n,0},\\
\label{fcommtwist}
\{ d^{\alpha A}_m, d^{\beta B}_n\}&=&-2\epsilon^{\alpha\beta}\epsilon^{AB}\delta_{m+n,0}.
\eea

We would like to find an expression for the state
\be\label{chionsigma}
|\chi(w_0)\rangle\equiv \sigma_2^+(w_0)|0_R^-\rangle^{(1)}\otimes |0_R^-\rangle^{(2)}
\ee
in terms of operator modes acting on $|0^-_R\rangle$. To do so, \cite{acm10a,acm10b} started from the initial state defined in the Ramond sector on the cylinder and then performed a series of spectral flow transformations and coordinate changes and map the state to a state of simpler form. We briefly describe the process below:

\b

($i$) The first step is to perform a spectral flow on the cylinder (\ref{spectral}) with parameter $\alpha=1$. This takes the two copies of the CFT from the Ramond sector to the NS sector. The bosons are not affected by the spectral flow, while the fermions transform as
\be\label{sf1bb}
\psi^{+A}(w)\r e^{-{\frac{w}{2}}}\psi^{+A}(w), ~~~\psi^{-A}(w)\r e^{{\frac{w}{2}}}\psi^{-A}(w).
\ee

\b

($ii$) We next map the cylinder with coordinate $w$ to the complex plane with coordinate $z$ via the map
\be
z=e^w.
\ee
The two NS vacua at $\tau\to-\infty$ on the cylinder are mapped into two NS vacua at $z=0$ on the complex plane. The location of the twist operator is mapped into $z_0=e^{w_0}$.

\begin{figure}[ht]
\begin{center}
\includegraphics[width=10cm]{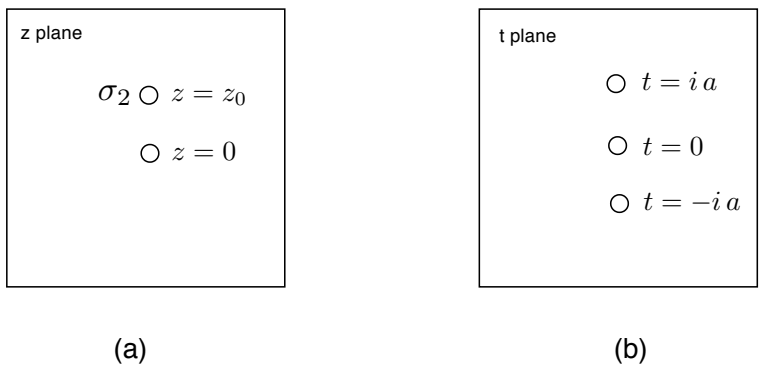}
\caption{The map from the complex plane to its covering surface is given by $z=z_0+t^2$, where $z$ is the coordinate on the complex plane and $t$ is the coordinate on the covering space. The two NS vacua at $z=0$ are mapped to two NS vacua at $t=\pm ia$ (these correspond the original two Ramond vacua at $\tau\to-\infty$ on the cylinder). The location of the twist insertion at $z=z_0$ is mapped to $t=0$ on the cover.
}\label{t-plane}
\end{center}
\end{figure}

\b

($iii$) We pass from the complex plane to its covering surface through the map
\be
z=z_0+t^2,
\ee
where
\be
z_0=e^{w_0}\equiv a^2.
\ee
This transformation is described in figure \ref{t-plane}. Passing to the covering space allows us to analyze explicitly the action of $\sigma^+_2$ on the initial state. The two NS vacua at $z=0$ in the $z$-plane are mapped into two NS vacua at the two punctures at $t=\pm ia$ in the $t$ plane. Since there are no operator insertions at these punctures, we can close them smoothly. The twist operator $\sigma_2^+$ at $z=z_0$ gets mapped into the spin up Ramond vacuum state of the covering space, $|0^+_R\rangle_t$, at $t=0$.

\b

($iv$) We finally perform a spectral flow with parameter $\alpha=-1$ on the covering space . This operation maps the Ramond vacuum state at $t=0$ to the NS vacuum in the $t$-space. There no other insertions at this point on the $t$-plane, so we can smoothly close the puncture at $t=0$. The bosons are not affected by the spectral flow transformation but the fermions transform as
\be\label{sf2bb}
\psi^{+A}\r t^\h\psi^{+A}, ~~~\psi^{-A}\r t^{-\h} \psi^{-A}.
\ee

\b

We thus find that the above sequence of spectral flow and coordinate transformations maps the initial state $|\chi\rangle$ in (\ref{chionsigma}) to the NS vacuum state in the $t$ plane at large $t$ (\emph{i.e.} $\tau\to\infty$). This vacuum state is then expressed in terms of the original modes on the cylinder at $\tau>\tau_0$. In order to do so, \cite{acm10a} computes how the bosonic and fermionic modes on the cylinder are transformed under the operations ($i$)-($iv$). At $\tau<\tau_0$ the $t$-plane modes are
\bea
\alpha^{(1)}_{A\dot A, n}&\r& {\frac{1}{2\pi}} \int_{t=ia} \p_t X_{A\dot A}(t) (z_0+t^2)^n dt,
\label{lseven}\\
\alpha^{(2)}_{A\dot A, n}&\r& {\frac{1}{2\pi}} \int_{t=-ia} \p_t X_{A\dot A}(t) (z_0+t^2)^n dt,
\label{leight}\\
d^{(1)+ A}_n&\r& {\frac{2^\h}{2\pi i}} \int_{t=ia} \psi^{+ A}(t) (z_0+t^2)^{n-1}t dt,
\label{lnine}\\
d^{(2)+ A}_n&\r& {\frac{2^\h}{2\pi i}} \int_{t=-ia} \psi^{+ A}(t) (z_0+t^2)^{n-1}t dt,
\label{lten}\\
d^{(1)- A}_n&\r&{\frac{2^\h}{2\pi i}} \int_{t=ia} \psi^{- A}(t) (z_0+t^2)^{n}  dt,
\label{lel}\\
d^{(2)- A}_n&\r&{\frac{2^\h}{2\pi i}} \int_{t=-ia} \psi^{- A}(t) (z_0+t^2)^{n}  dt,
\label{ltw}
\eea
and at $\tau>\tau_0$ the modes read
\bea
\alpha_{A\dot A, n}&\r& {\frac{1}{2\pi}} \int_{t=\infty} \p_t X_{A\dot A}(t) (z_0+t^2)^{{\frac{n}{2}}} dt,
\label{qatwo}\\
d^{+ A}_n&\r& {\frac{2^\h}{2\pi i}}\int_{t=\infty} \psi^{+ A}(t) (z_0+t^2)^{{\frac{(n-2)}{2} }} t dt,
\label{lthir}\\
d^{- A}_n&\r& {\frac{2^\h}{2\pi i}}\int_{t=\infty} \psi^{- A}(t) (z_0+t^2)^{{\frac{n}{2}} }  dt\label{lthirf}.
\eea

In the $t$ plane, we will also define operator modes that are natural to the $t$ plane. Thus we write
\be
\t\alpha_{A\dot A, n}\equiv  {\frac{1}{ 2\pi}} \int_{t=0} \p_t X_{A\dot A}(t) t^n dt,
\label{qathree}
\ee
\be
\t d^{\alpha A}_r= {\frac{1}{2\pi i}}\int_{t=0} \psi^{\alpha A}(t) t^{r-\h} dt.
\ee
Note that the bosonic index $n$ is an integer while the fermionic index $r$ is a half integer.  We have
\be
\t\alpha_{A \dot A,m}|0\rangle_t=0, ~~~m\ge 0,
\label{ptthree}
\ee
\be
\t d^{\alpha A}_r|0\rangle_t=0, ~~~r>0.
\label{ptthreep}
\ee

The computation of \cite{acm10a} found that the state $|\chi\rangle$ has the structure of a squeezed state
\be
|\chi\rangle=e^{\sum_{ m\ge 1, n\ge 1}\gamma^B_{mn}[-\alpha_{++, -m}\alpha_{--, -n}+\alpha_{-+, -m}\alpha_{+-, -n}]}\;
e^{\sum_{m\ge0,n\ge 1}\gamma^F_{mn}[d^{++}_{-m}d^{--}_{-n}-d^{+-}_{-m}d^{-+}_{-n}]}|0^-_R\rangle,
\label{pfive}
\ee
where the coefficients $\gamma^B_{m,n}$ and $\gamma^F_{m,n}$ are given by
\be\label{gammaB}
\gamma^B_{2m'+1, 2n'+1}={\frac{2}{(2m'+1)(2n'+1)}} \frac{z_0^{(1+m'+n')} \Gamma[{\frac{3}{2}}+m']\Gamma[{\frac{3}{2}}+n']}{(1+m'+n')\pi \Gamma[m'+1]\Gamma[n'+1]},
\ee
\be\label{gammaF}
\gamma^{F}_{2m'+1, 2n'+1}=-\frac{ z_0^{(1+m'+n')}\Gamma[{\frac{3}{2}}+m']\Gamma[{\frac{3}{2}}+n']}{(2n'+1)\pi(1+m'+n') \gamma[m'+1]\Gamma[n'+1]}.
\ee
The full state (\ref{fullstate}) is then obtained by the action of the supercharge on the squeezed state. We refer the reader to \cite{acm10a} for the details of the computation of the full state.

\section{Action of the stress tensor}\label{sectionthree}

Consider the cylinder shown in fig. \ref{2-twist}, with state $|0_R^-\rangle^{(1)}\otimes |0_R^-\rangle^{(2)}$ at $\tau\rightarrow-\infty$ and the twist $\sigma_2^+$ inserted at $w_0$. Let $T$ be the stress tensor of the CFT, given in  (\ref{Top}). We integrate $T$ around a small contour around $w_0$, getting
\be
\oint_{w=w_0} \frac{dw}{2\pi i}\;T(w)\;\sigma_2^+(w_0)|0_R^-\rangle^{(1)}\otimes |0_R^-\rangle^{(2)}=\partial_{w_0}\sigma_2^+(w_0)|0_R^-\rangle^{(1)}\otimes |0_R^-\rangle^{(2)}=\partial_{w_0}|\chi(w_0)\rangle
\ee
where we have used the fact that $\sigma_2^+$ is a primary of the Virasoro algebra. Thus if we can understand the action of $T$ in some other description, then we can arrive at a description of $\partial_{w_0}|\chi(w_0)\rangle$, and by integration, at a description for $|\chi(w_0)\rangle$ itself. This then gives a simple method to evaluate the coefficients $\gamma^B_{mn}$ and $\gamma^F_{mn}$ by only using basic properties of the CFT.

\subsection{Computation of \texorpdfstring{$\partial_{w_0}|\chi(w_0)\rangle$}{Lg}}\label{iii-i}

To get this alternative description, we follow the steps in subsection (\ref{ii-i}).

\b
(i) We first start from the contour integral on the cylinder and consider a description which is spectral flowed by $\alpha=1$. We obtain
\begin{eqnarray}\label{Tsigma2-2}
&&\oint_{w=w_0}\frac{dw}{2\pi i}\sum_{i=1}^{2}T(w)^{(i)}
\sigma^+_2(w_0)|0^-_R\rangle^{(1)}\otimes|0^-_R\rangle^{(2)}\xrightarrow{sf_1}\nonumber\\
&&\qquad\oint_{w=w_0}\frac{dw}{2\pi i}\sum_{i=1}^{2}\Big(T^{(i)}(w)-J^{(i)\,3}(w)+\frac14\Big)\,
e^{-\frac12w_0}\sigma^+_2(w_0)|0\rangle^{(1)}\otimes|0\rangle^{(2)},
\end{eqnarray}
where $T(w)=\sum_{i=1}^{2}T^{(i)}(w)$ and $\xrightarrow{sf_1}$ represents the spectral flow with $\alpha=1$. We note that the spectral flow with $\alpha=1$ maps the Ramond vacua $|0^-_R\rangle^{(i)}$ to NS vacua $|0\rangle^{(i)}$ and acts on the chiral primary as
\begin{equation}\label{sf-w-sigma2}
\sigma_2^+(w_0)\to e^{-\frac12w_0}\sigma_2^+.
\end{equation}
The transformation of the stress-energy tensor under the spectral flow is derived in appendix \ref{appb} and is given by
 \begin{eqnarray}\label{sf-w-T-8}
T^{(i)}(w)\xrightarrow{sf_1}\;T^{(i)}(w)-J^{(i)\,3}(w)+\frac14.
\label{aten}
\end{eqnarray}
The integral of the constant term ($1/4$) on the right-hand-side of (\ref{Tsigma2-2}) vanishes and thus we find
\bea\label{Tsigma2-2-ii}
&&\oint_{w_0}\frac{dw}{2\pi i}\;T(w)\sigma_2^+|0^-_R\rangle^{(1)}\otimes|0^-_R\rangle^{(2)}\xrightarrow{sf_1}\nonumber\\
&&\qquad\oint_{w_0}\frac{dw}{2\pi i}\sum_{i=1}^{2}\Big(T^{(i)}(w)-J^{(i)\,3}(w)\Big)\,
e^{-\frac12w_0}\sigma^+_2(w_0)|0\rangle^{(1)}\otimes|0\rangle^{(2)}.
\eea

\b

(ii) Next, we perform coordinate transformation $z=e^w$ to map $T(w)$ to the complex plane. The NS vacua at $\tau\to-\infty$ are mapped into vacua at $z=0$ and the puncture is smoothly closed since there are no insertions at this point. We then consider how $T(w)$, $J(w)$, and $\sigma_2^+(w)$ transform under this map. The stress tensor is a quasi-primary operator and is modified by the Schwarzian derivative under the coordinate transformation:
\begin{equation}\label{T-map-wz-1}
\left(\frac{\partial z^\prime}{\partial z}\right)^2T^{(i)}(z^\prime)=T^{(i)}(z)-\frac{c}{12}\{z^\prime,z\},
\end{equation}
where $\{z^\prime,z\}$ is the Schwarzian derivative
\begin{equation}\label{schwzn}
\{z^\prime,z\}=\frac{\frac{\partial z^\prime}{\partial z}\,\frac{\partial^3 z^\prime}{{\partial z}^3}-
\frac32\left(\frac{\partial^2 z^\prime}{{\partial z}^2}\right)^2}
{\left(\frac{\partial z^\prime}{\partial z}\right)^2}.
\end{equation}
Using (\ref{schwzn}), equation (\ref{T-map-wz-1}) reads
\begin{equation}\label{T-map-wz-2}
T^{(i)}(w)=z^2\,T^{(i)}(z)-\frac{1}{4}.
\end{equation}
The $R$-current and the chiral primary twist operator transform as $J^{(i)}(w)=z\,J^{(i)}(z)$ and $\sigma_2^+(w_0)=e^{\frac12w_0}\sigma_2^+(z_0)$, and we find
\begin{eqnarray}\label{Tsigma2-3}
&&\oint_{w_0}\frac{dw}{2\pi i}\sum_{i=1}^{2}\Big(T^{(i)}(w)-J^{(i)\,3}(w)\Big)\,
e^{-\frac12w_0}\sigma^+_2(w_0)|0\rangle^{(1)}\otimes|0\rangle^{(2)}\to\nonumber\\
&&\qquad\oint_{z_0}\frac{dz}{2\pi i}\;\frac{\partial w}{\partial z}\sum_{i=1}^{2}
\Big(z^2\,T^{(i)}(z)-\frac14-zJ^{(i)\,3}(z)\Big)\,\sigma^+_2(z_0)|0\rangle^{(1)}\otimes|0\rangle^{(2)}.
\end{eqnarray}

\b

(iii) We next pass from the complex plane to the covering surface through the map $z=z_0+t^2$.
The two initial NS vacua are mapped into punctures at $t=\pm ia$ ($a=z_0^{1/2}$).
There are no other insertions at these two punctures and one can smoothly close them.
The chiral primary at $z=z_0$ is mapped into a puncture at $t=0$ at which we have the positive spin Ramond vacuum $|0_R^+\rangle_t$.
Under this coordinate transformation we have
\begin{equation}\label{T-map-zt-1}
T(z)=\frac{1}{4t^2}T(t)+\frac{3}{16t^4},\qquad\qquad J(z)=\frac{1}{2t}J(t),
\end{equation}
and the integral reads
\begin{eqnarray}\label{Tsigma2-4}
&&\oint_{z_0}\frac{dz}{2\pi i}\;\frac{1}{z}\;\sum_{i=1}^{2}
\Big(z^2\,T^{(i)}(z)-\frac14-zJ^{(i)\,3}(z)\Big)\,\sigma^+_2(z_0)|0\rangle^{(1)}\otimes|0\rangle^{(2)}\to\nonumber\\
&&\qquad\qquad\oint_{t=0}\frac{dt}{2\pi i}\;\frac{1}{z}\;\frac{\partial z}{\partial t}\,
\bigg(z^2\Big(\frac{1}{4t^2}T(t)+\frac{3}{16t^4}\Big)-\frac14-\frac{z}{2t}J^{3}(t)\bigg)\,|0^+_R\rangle_t=\nonumber\\
&&\qquad\qquad\oint_{t=0}\frac{dt}{2\pi i}\,\bigg(\frac{z}{2t}\,T(t)-J^{3}(t)+\frac{3z}{8t^3}-\frac{t}{2z}\bigg)
\,|0^+_R\rangle_t,
\end{eqnarray}
where the integral of the constant term ($-1/4$) in the second line vanishes. We further note that
\begin{equation}\label{Tsigma2-4-i}
\oint_{t=0}\frac{dt}{2\pi i}\;\frac{t}{2z}\;|0^+_R\rangle_t=0,
\end{equation}
and
\begin{equation}\label{Tsigma2-4-ii}
\oint_{t=0}\frac{dt}{2\pi i}\;\frac{3\,z}{8t^3}\;|0^+_R\rangle_t=
\oint_{t=0}\frac{dt}{2\pi i}\;\frac{3\,(z_0+t^2)}{8t^3}\,|0^+_R\rangle_t=
\frac{3}{8}\,|0^+_R\rangle_t.
\end{equation}
Using the fact that $|0^+_R\rangle_t$ has $R$-charge $+1/2$, we find
\begin{equation}\label{Tsigma2-4-iii}
\frac{1}{2\pi i}\oint_{t=0}dtJ^3(t)\,|0^+_R\rangle_t=\frac{1}{2}\,|0^+_R\rangle_t.
\end{equation}
Equation (\ref{Tsigma2-4}) then reads
\begin{eqnarray}\label{Tsigma2-5}
&&\oint_{z_0}\frac{dz}{2\pi i}\;\frac1{z}\;\sum_{i=1}^{2}\Big(z^2\,T^{(i)}(z)-\frac14-zJ^{(i)\,3}(z)\Big)\,\sigma^+_2(z_0)\to\nonumber\\
&&\qquad\qquad\qquad\oint_{t=0}\frac{dt}{2\pi i}\,\frac{(z_0+t^2)}{2t}\,T(t)\,|0^+_R\rangle_t-\frac18\,|0^+_R\rangle_t.
\end{eqnarray}

\b

(iv) Finally, we perform a spectral flow in the $t$ plane with spectral flow parameter $\alpha=-1$. This maps the Ramond vacuum at $t=0$ to a NS vacuum and we can then smoothly close the puncture at this point on the cover. The transformation of $T$ under this spectral flow is of the form (see appendix \ref{appb})
\bea\label{sf-t-T-6}
T(t)\xrightarrow{sf_2}T(t)+\frac{1}{t}J^{3}(t)+\frac{1}{4t^2},
\label{sone}
\eea
where $\xrightarrow{sf_2}$ corresponds to the spectral flow on the covering surface with $\alpha=-1$. We thus find that
\begin{eqnarray}\label{Tsigma2-6}
&&\oint_{t=0}\frac{dt}{2\pi i}\;\frac{(z_0+t^2)}{2t}\;T(t)\;|0^+_R\rangle_t-\frac18\;|0^+_R\rangle_t\xrightarrow{sf_2}\nonumber\\
&&\qquad\qquad\oint_{t=0}\frac{dt}{2\pi i}\;\frac{(z_0+t^2)}{2t}\;\Big(T(t)+\frac{1}{t}J^3(t)+\frac{1}{4t^2}\Big)\;|0\rangle_t-\frac18\;|0\rangle_t.
\end{eqnarray}
Let us consider the last term inside the parentheses in the right-hand-side of the above equation. The contour integral for this term reads
\begin{equation}\label{Tsigma2-4-i-ii}
\oint_{t=0}\frac{dt}{2\pi i}\;\frac{(z_0+t^2)}{8t^3}\;|0\rangle_t=\frac18\;|0\rangle_t.
\end{equation}
Inserting this in equation (\ref{Tsigma2-6}), the constant parts cancel out and we obtain
\begin{eqnarray}\label{Tsigma2-7-i}
&&\oint_{t=0}\frac{dt}{2\pi i}\;\frac{(z_0+t^2)}{2t}\;T(t)\;|0^+_R\rangle_t-\frac18\;|0^+_R\rangle_t\xrightarrow{sf_2}\nonumber\\
&&\qquad\oint_{t=0}\frac{dt}{2\pi i}\;\frac{(z_0+t^2)}{2t}\;\Big(T(t)+\frac{1}{t}J^3(t)\Big)\;|0\rangle_t\,|0\rangle_t=
\frac{z_0}{2}\;\Big(\tilde L_{-2}+\tilde J^3_{-2}\Big)\,|0\rangle_t,
\end{eqnarray}
where $\tilde L_m$ and $\tilde J^3_m$ are the modes natural to the covering surface given by
\begin{eqnarray}\label{TJ-t-mode}
\tilde L_n&=&\oint \frac{dt}{2\pi i}\;t^{n+1}\;T(t),\\
\tilde J^3_n&=&\oint \frac{dt}{2\pi i}\;t^{n}\;J^3(t).
\end{eqnarray}

We thus find that under the series of maps and spectral flows explained above the action of $T$ on $\sigma_2^+$ is given by
\begin{eqnarray}\label{Tsigma2-7}
&&\oint_{w_0}\frac{dw}{2\pi i}\;T(w)\;\sigma^+_2(w_0)|0^-_R\rangle^{(1)}\otimes|0^-_R\rangle^{(2)}=\partial_{w_0}|\chi(w_0)\rangle\longrightarrow
\frac{z_0}{2}\;\Big(\tilde L_{-2}+\tilde J^3_{-2}\Big)\,|0\rangle_t.
\end{eqnarray}


\subsection{Writing \texorpdfstring{$\partial_{w_0}|\chi(w_0)\rangle$}{Lg} in terms of boson and fermion modes on the cover}

We now write the above result (\ref{Tsigma2-7}) in terms of boson and fermion modes which are natural to the $t$ plane. Modes of bosonic fields are
\begin{eqnarray}\label{delX-mode-t}
\partial_tX_{A\dot A}=-i\sum_{m\in\mathbb{Z}}\frac{\tilde\alpha_{A\dot A,m}}{t^{m+1}},
\end{eqnarray}
and modes of fermionic fields in the NS sector are
\begin{eqnarray}\label{psi-mode-t}
\psi^{\alpha A}=\sum_{r\in\mathbb{Z}+\frac12}\frac{\tilde d^{\alpha A}_r}{t^{r+\frac12}}.
\end{eqnarray}
These give
\begin{equation}\label{delX-t-0}
\tilde\alpha_{A\dot A,m}=\frac{1}{2\pi}\oint_{t=0}dt\,t^{m}\,\partial_tX_{A\dot A}(t),
\quad n\in\mathbb{Z},
\end{equation}
\begin{equation}\label{psi-t-0}
\tilde d^{\alpha A}_{r}=\frac{1}{2\pi i}\oint_{t=0}dt\,t^{r-\frac12}\,\psi^{\alpha A}(t),
\quad r\in\mathbb{Z}+\frac12.
\end{equation}
Inserting these expansions in the stress-energy tensor (\ref{Top}) and the $R$-current (\ref{Jop}) we expand the generators $\tilde L_n$ and $\tilde J^3_n$ in terms of $\tilde\alpha_{A\dot A,m}$ and $\tilde d^{\alpha A}$:
\begin{equation}\label{Ln-t}
\tilde L_n=\sum_{m\in\mathbb Z}(-\tilde\alpha_{+\dot+,n-m}\,\tilde\alpha_{-\dot-,m}+
\tilde\alpha_{+\dot-,n-m}\,\tilde\alpha_{-\dot+,m})
+\sum_{r\in\mathbb Z+\frac12}(\frac{n}{2}-r)\,(\tilde d^{++}_{n-r}\,\tilde d^{--}_{r}-
\tilde d^{+-}_{n-r}\,\tilde d^{-+}_{r}),
\end{equation}
and
\begin{equation}\label{Jn-t}
\tilde J^3_n=\frac12\sum_{r\in\mathbb Z+\frac12}(-\tilde d^{++}_{n-r}\,\tilde d^{--}_{r}+
\tilde d^{+-}_{n-r}\,\tilde d^{-+}_{r}).
\end{equation}
For $n=-2$ we find that
\begin{eqnarray}\label{L-2-0}
\tilde L_{-2}\,|0\rangle_t&=&
(-\tilde\alpha_{+\dot+,-1}\,\tilde\alpha_{-\dot-,-1}+
\tilde\alpha_{+\dot-,-1}\,\tilde\alpha_{-\dot+,-1})\,|0\rangle_t\nonumber\\
&+&\frac12\,(\tilde d^{++}_{-\frac12}\,\tilde d^{--}_{-\frac32}-
\tilde d^{++}_{-\frac32}\,\tilde d^{--}_{-\frac12}
-\tilde d^{+-}_{-\frac12}\,\tilde d^{-+}_{-\frac32}
+\tilde d^{+-}_{-\frac32}\,\tilde d^{-+}_{-\frac12})\,|0\rangle_t,
\end{eqnarray}
\begin{equation}\label{J-2-0}\tilde J^3_{-2}\,|0\rangle_t=\frac12(
-\tilde d^{++}_{-\frac12}\,\tilde d^{--}_{-\frac32}
-\tilde d^{++}_{-\frac32}\,\tilde d^{--}_{-\frac12}
+\tilde d^{+-}_{-\frac12}\,\tilde d^{-+}_{-\frac32}
+\tilde d^{+-}_{-\frac32}\,\tilde d^{-+}_{-\frac12})\,|0\rangle_t.
\end{equation}
Adding together the above two equations we obtain
\begin{equation}\label{L-2+J-2-0}
(\tilde L_{-2}+\tilde J^3_{-2})\,|0\rangle_t=
\Big(-\tilde\alpha_{+\dot+,-1}\,\tilde\alpha_{-\dot-,-1}+
\tilde\alpha_{+\dot-,-1}\,\tilde\alpha_{-\dot+,-1}
-\tilde d^{++}_{-\frac32}\,\tilde d^{--}_{-\frac12}
+\tilde d^{+-}_{-\frac32}\,\tilde d^{-+}_{-\frac12}\Big)|0\rangle_t.
\end{equation}
Inserting this expression back in (\ref{Tsigma2-7}), the action of $T$ on $|\chi\rangle$ is given by
\begin{equation}\label{Tsigma2-8}
\partial_{w_0}|\chi(w_0)\rangle\to\frac{z_0}{2}\,\Big(-\tilde\alpha_{+\dot+,-1}\,\tilde\alpha_{-\dot-,-1}+
\tilde\alpha_{+\dot-,-1}\,\tilde\alpha_{-\dot+,-1}
-\tilde d^{++}_{-\frac32}\,\tilde d^{--}_{-\frac12}
+\tilde d^{+-}_{-\frac32}\,\tilde d^{-+}_{-\frac12}\Big)\,|0\rangle_t.
\end{equation}

\subsection{Computing the coefficients \texorpdfstring{$\gamma^B_{mn}$}{Lg} and \texorpdfstring{$\gamma^F_{mn}$}{Lg}}
We express the above result in terms of modes on the cylinder.  The modes $\tilde\alpha_{A\dot A,n}$ (\ref{delX-t-0}) and $\tilde d^{\alpha A}_{r}$ (\ref{psi-t-0}) where defined by contour integrals over circles around $t=0$.
Since there are no singularities at any point in the $t$ plane, we can stretch the contours to circles at large $t$
\begin{equation}\label{delX-t-infty-1}
\tilde\alpha_{A\dot A,m}=\frac{1}{2\pi}\int_{t=\infty}dt\,t^{m}\,\partial_tX_{A\dot A},
\quad m\in\mathbb{Z},
\end{equation}
\begin{equation}\label{psi-t-infty-1}
\tilde d^{\alpha A}_{r}=\frac{1}{2\pi i}\int_{t=\infty}dt\,t^{r-\frac12}\,\psi^{\alpha A},
\quad r\in\mathbb{Z}+\frac12.
\end{equation}
We can now convert these modes to the modes on the cylinder at $\tau\to\infty$ derived in section \ref{sectiontwo}. Under the spectral flows and coordinate maps that we explained above, the modes on the cylinder at $\tau\to\infty$ transform into (\ref{qatwo}), (\ref{lthir}), and (\ref{lthirf}).
We write
\begin{eqnarray}\label{t-bos-xpn}
t^m&=&(t^2+z_0-z_0)^{\frac{m}{2}}=(z_0+t^2)^{\frac{m}{2}}
\Big(1-z_0\,(z_0+t^2)^{-1}\Big)^{\frac{m}{2}}\nonumber\\
&=&\sum_{k\ge0}\,^{\frac m2}C_k\,(-z_0)^k\,(z_0+t^2)^{\frac m2-k},
\end{eqnarray}
where $^qC_k= \binom{q}{k}$ is the binomial coefficient.
Substituting this in the mode expansion (\ref{delX-t-infty-1}) we have
\begin{eqnarray}\label{delX-t-infty-2}
\tilde\alpha_{A\dot A,m}&=&\frac{1}{2\pi}\int_{t=\infty}dt\,
\sum_{k\ge0}\,^{\frac m2}C_k\,(-z_0)^k
\,\partial_tX_{A\dot A}(z_0+t^2)^{\frac {m-2k}2}\nonumber\\
&=&\sum_{k\ge0}\,^{\frac m2}C_k\,(-z_0)^k\,\alpha_{A\dot A,m-2k},
\end{eqnarray}
where we used (\ref{qatwo}) to get to the second line. Similarly, we write
\begin{equation}\label{t-ferp-xpn}
t^{r-\frac12}=t\,(t^2+z_0-z_0)^{\frac{1}{2}(r-\frac32)}=
t\,\sum_{k\ge0}\,^{\frac12(r-\frac32)}C_k\,(-z_0)^k\,(z_0+t^2)^{\frac12(r-\frac32)-k}.
\end{equation}
Substituting this in equation (\ref{psi-t-infty-1}) we find that
\begin{eqnarray}\label{psi-t-infty-2}
\tilde d^{+A}_{r}&=&\frac{1}{2\pi i}\int_{t=\infty}dt
\sum_{k\ge0}\,^{\frac12(r-\frac32)}C_k\,(-z_0)^k
\,\psi^{+A},(z_0+t^2)^{\frac{r-2k-\frac32}{2}}\,t\nonumber\\
&=&2^{-\frac12}\sum_{k\ge0}\,^{\frac12(r-\frac32)}C_k\,(-z_0)^k\,d^{+A}_{r-2k+\frac12},
\end{eqnarray}
where we used (\ref{lthir}) in the second line. To find the negatively charged fermionic modes we write
\begin{equation}\label{t-ferm-xpn}
t^{r-\frac12}=(t^2+z_0-z_0)^{\frac{1}{2}(r-\frac12)}=
\sum_{k\ge0}\,^{\frac12(r-\frac12)}C_k\,(-z_0)^k\,(z_0+t^2)^{\frac12(r-\frac12)-k}.
\end{equation}
Substituting this in equation (\ref{psi-t-infty-1}) and using (\ref{lthirf}) we have
\begin{eqnarray}\label{psi-t-infty-3}
\tilde d^{-A}_{r}&=&\frac{1}{2\pi i}\int_{t=\infty}dt
\sum_{k\ge0}\,^{\frac12(r-\frac12)}C_k\,(-z_0)^k
\,\psi^{-A},(z_0+t^2)^{\frac{r-2k-\frac12}{2}}\,t\nonumber\\
&=&2^{-\frac12}\sum_{k\ge0}\,^{\frac12(r-\frac12)}C_k\,(-z_0)^k\,d^{-A}_{r-2k-\frac12}.
\end{eqnarray}
We finally express (\ref{Tsigma2-8}) in terms of the operator modes on the cylinder
\begin{eqnarray}\label{Tsigma2-9}
&&\!\!\!\!\!\!\!\!\!\!\!\!\!\!\!\!\!\!\!\!\partial_{w_0}|\chi(w_0)\rangle\to\nonumber\\
\bigg[\!\!\!&-&z_0\,\Big(\frac{1}{\sqrt2}\sum_{k\ge0}\,^{-\frac{1}2}C_k
\,(-z_0)^k\,\alpha_{+\dot+,-1-2k}\Big)
\Big(\frac{1}{\sqrt2}\sum_{k^\prime\ge0}\,^{-\frac{1}2}C_{k^\prime}
\,(-z_0)^{k^\prime}\,\alpha_{-\dot-,-1-2k^\prime}\Big)\nonumber\\
&+&z_0\,\Big(\frac{1}{\sqrt2}\sum_{k\ge0}\,^{-\frac{1}2}C_k
\,(-z_0)^k\,\alpha_{+\dot-,-1-2k}\Big)
\Big(\frac{1}{\sqrt2}\sum_{k^\prime\ge0}\,^{-\frac{1}2}C_{k^\prime}
\,(-z_0)^{k^\prime}\,\alpha_{-\dot+,-1-2k^\prime}\Big)\nonumber\\
&-&z_0\Big(\frac12\sum_{k\ge0}\,^{-\frac32}C_k
\,(-z_0)^k\,d^{++}_{-1-2k}\Big)
\Big(\frac12\sum_{k^\prime\ge0}\,^{-\frac12}C_{k^\prime}
\,(-z_0)^{k^\prime}\,d^{--}_{-1-2k^\prime}\Big)\nonumber\\
&+&z_0\Big(\frac12\sum_{k\ge0}\,^{-\frac32}C_k
\,(-z_0)^k\,d^{+-}_{-1-2k}\Big)
\Big(\frac12\sum_{k^\prime\ge0}\,^{-\frac12}C_{k^\prime}
\,(-z_0)^{k^\prime}\,d^{-+}_{-1-2k^\prime}\Big)\bigg]\,|0\rangle_t.
\end{eqnarray}

We evaluate the coefficients $\gamma^B_{mn}$ and $\gamma^F_{mn}$ by integrating the above equation. As explained in section \ref{sectiontwo}, the state $|\chi(w_0)\rangle=\sigma^+_2(w_0)|0^-_R\rangle^{(1)}\otimes|0^-_R\rangle^{(2)}$ is mapped into the $t$ plane NS vacuum under the spectral flows and coordinate transformations that we apply. It was shown in \cite{acm10a} that  $|\chi(w_0)\rangle$ has the form of a squeezed state (\ref{pfive}). We thus obtain that for this squeezed state
\begin{eqnarray}\label{dsigmadw0}
&&\!\!\!\!\!\!\!\!\!\!\!\!\!\!\!\partial_{w_0}|\chi(w_0)\rangle=\nonumber\\
&\Bigg(&\!\!\sum_{m\ge1,n\ge1}\partial_{w_0}\gamma^B_{mn}
\left(-\alpha_{++,-m}\alpha_{--,-n}+\alpha_{-+,-m}\alpha_{+-,-n} \right)\nonumber\\
&+&\sum_{m\ge0,n\ge1}\partial_{w_0}\gamma^F_{mn}
\left(d^{++}_{-m}d^{--}_{-n}-d^{+-}_{-m}d^{-+}_{-n} \right)\Bigg)
\sigma_2^+(w_0)|0^-_R\rangle^{(1)}\otimes|0^-_R\rangle^{(2)}.
\end{eqnarray}
Comparing (\ref{Tsigma2-9}) and (\ref{dsigmadw0}) we find that for $m$ and $n$ being odd and positive integers $m=2m^\prime+1$, $m^\prime\ge0$ and $n=2n^\prime+1$, $n^\prime\ge0$, we have
\begin{eqnarray}\label{gamma-bos-1}
\partial_{w_0}\gamma^B_{2m^\prime+1,2n^\prime+1}=z_0\,\partial_{z_0}\gamma^B_{2m^\prime+1,2n^\prime+1}=
z_0\,\Big(\frac{1}{\sqrt2}\,^{-\frac{1}2}C_{m^\prime}\,(-z_0)^{m^\prime}\Big)
\Big(\frac{1}{\sqrt2}\,^{-\frac{1}2}C_{n^\prime}\,(-z_0)^{n^\prime}\Big),
\end{eqnarray}
\begin{eqnarray}\label{gamma-fer-1}
\partial_{w_0}\gamma^F_{2m^\prime+1,2n^\prime+1}=z_0\,\partial_{z_0}\gamma^F_{2m^\prime+1,2n^\prime+1}=
-z_0\,\Big(\frac1{2\sqrt2}\,^{-\frac{3}2}C_{m^\prime}\,(-z_0)^{m^\prime}\Big)
\Big(\frac{1}{\sqrt2}\,^{-\frac{1}2}C_{n^\prime}\,(-z_0)^{n^\prime}\Big).
\end{eqnarray}
This then gives
\begin{equation}\label{gamma-bos-2}
z_0\,\partial_{z_0}\gamma^B_{2m^\prime+1,2n^\prime+1}=z_0\,\Bigg(\frac{\sqrt2}{\sqrt\pi}\frac{z_0^{m^\prime}}{(2m^\prime+1)}\,\frac{\Gamma[\frac32+m^\prime]}{\Gamma[1+m^\prime]}\Bigg)
\Bigg(\frac{\sqrt2}{\sqrt\pi}\frac{z_0^{n^\prime}}{(2n^\prime+1)}\,\frac{\Gamma[\frac32+n^\prime]}{\Gamma[1+n^\prime]}\Bigg),
\end{equation}
and
\begin{equation}\label{gamma-fer-2}
z_0\,\partial_{z_0}\gamma^F_{2m^\prime+1,2n^\prime+1}=-z_0\,\Bigg(\frac{1}{\sqrt{2\pi}}\,{z_0^{m^\prime}}\,\frac{\Gamma[\frac32+m^\prime]}{\Gamma[1+m^\prime]}\Bigg)
\Bigg(\frac{\sqrt2}{\sqrt\pi}\frac{z_0^{n^\prime}}{(2n^\prime+1)}\,\frac{\Gamma[\frac32+n^\prime]}{\Gamma[1+n^\prime]}\Bigg).
\end{equation}
We observe that the derivative of $|\chi\rangle$ is given in terms of a product of two factors $f(m)$ and $f(n)$.

Using (\ref{gamma-bos-2}) and (\ref{gamma-fer-2}), we integrate (\ref{dsigmadw0}) with respect to $z_0$ and find that
\begin{eqnarray}\label{chichi-squeez}
&&|\chi(w_0)\rangle=\sigma_2^+(w_0)\,|0^-_R\rangle^{(1)}\otimes|0^-_R\rangle^{(2)}=\\
&&e^{\sum_{m\ge1,n\ge1}\gamma^B_{mn}
\left(-\alpha_{++,-m}\alpha_{--,-n}+\alpha_{-+,-m}\alpha_{+-,-n} \right)}\,
e^{\sum_{m\ge0,n\ge1}\gamma^F_{mn}
\left(d^{++}_{-m}d^{--}_{-n}-d^{+-}_{-m}d^{-+}_{-n} \right)}|0^-_R\rangle\nonumber,
\end{eqnarray}
where
\begin{eqnarray}
\gamma^B_{2m^\prime+1,2n^\prime+1}&=&
\frac{2\,z_0^{(1+m^\prime+n^\prime)}}
{\pi\,(2m^\prime+1)(2n^\prime+1)(1+m^\prime+n^\prime)}\;
\frac{\Gamma[\frac32+m^\prime]\,\Gamma[\frac32+n^\prime]}
{\Gamma[1+m^\prime]\,\Gamma[1+n^\prime]},\label{gammab-ii}\\
\gamma^F_{2m^\prime+1,2n^\prime+1}&=&
-\frac{z_0^{(1+m^\prime+n^\prime)}}{\pi\,(2n^\prime+1)(1+m^\prime+n^\prime)}\;
\frac{\Gamma[\frac32+m^\prime]\,\Gamma[\frac32+n^\prime]}
{\Gamma[1+m^\prime]\,\Gamma[1+n^\prime]}.\label{gammaf-ii}
\end{eqnarray}
The result agrees with the coefficients $\gamma^B_{2m^\prime+1,2n^\prime+1}$ (\ref{gammaB}) and $\gamma^F_{2m^\prime+1,2n^\prime+1}$ (\ref{gammaF}) previously computed in \cite{acm10a}.

\section{Continuum limit}\label{sectionfour}
The D1-D5 system is constructed by wrapping the D1 branes on the circle $S^1$ and the D5 branes on $S^1\times T^4$. In the regime where the volume of the compactification circle is larger than the size of the torus, the D1-D5 CFT is a 2-dimensional CFT which lives on $S^1$. In the previous section we studied the action of the deformation operator on the vacuum state of the CFT. The deformation operator was inserted at the point $w_0=\tau_0+i\sigma_0$ on the cylinder corresponding to the 2-dimensional CFT.

\begin{figure}[ht]
\begin{center}
\includegraphics[width=8cm]{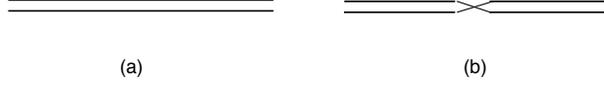}
\caption{The circles becomes infinite lines in the continuum limit, but the twist operator still works locally as it did in the compact case.}\label{contlimitbb}
\end{center}
\end{figure}

In this section we would like to study the behaviour of the final state in a region close to the insertion point of the twist operator. In this limit, the CFT lives on the non-compact infinite line $\mathcal{R}$ rather than the circle $S^1$ (see figure \ref{contlimitbb}). We refer to this limit as the continuum limit.

There are several places where this `local' description of the twist can be useful. First, if we are studying the CFT dual to the Poincare patch of $AdS_3\times S^3\times T^4$, then the CFT lives on ${\mathcal R}$ instead of $S^1$. Second, even if we have an $S^1$ compactification, the generic state dual to a black hole has a highly twisted sector of the orbifold CFT, thus creating a CFT that lives on an effectively infinite line. Third, when we are at strong coupling we expect that the quanta created by on twist insertion will be picked up and modified by another twist insertion before they have had time to travel around the closed loop on which the CFT lives; thus the CFT would again behave as if it lived on an infinite line.

In this section we will evaluate expressions describing the squeezed state in the continuum limit. These expressions enable us to better understand the physics of the excitations on top of the vacuum state created by the deformation operator as one deforms the CFT away from the orbifold point.  We will first analyze the bosonic part of the final state in section \ref{bcont-i}, and then consider the fermionic part in section \ref{fcont-i} . We will finally write down the full expression in the continuum limit in section \ref{cont-tot}.  These computations can be verified by directly exploring the squeezed state (\ref{pfive}) in the continuum limit, which we do in appendix \ref{altkernels}.
\subsection{Bosonic part of the squeezed state in the continuum limit}\label{bcont-i}
We consider again the expression that we obtained in section (\ref{sectionthree}) for the stress energy tensor method
\begin{eqnarray}
\!\!\!\!\!\!\!\!&&\frac{1}{2\pi i}\oint_{w_0}dwT(w)
\sigma^+_2(w_0)|0^-_R\rangle^{(1)}\otimes|0^-_R\rangle^{(2)}
=\partial_{w_0}|\chi\rangle=
\partial_{w_0}\sigma_2^+(w_0)|0^-_R\rangle^{(1)}\otimes|0^-_R\rangle^{(2)}\qquad \label{dsigma-2}\\
&&\to\;\frac{z_0}{2}\,\Big(\tilde{L}_{-2}+\tilde{J}^3_{-2}\Big)\,|0\rangle_t\nonumber\\
&&=\frac{z_0}{2}\,\Big(-\tilde\alpha_{+\dot+,-1}\,\tilde\alpha_{-\dot-,-1}+
\tilde\alpha_{+\dot-,-1}\,\tilde\alpha_{-\dot+,-1}
-\tilde d^{++}_{-\frac32}\,\tilde d^{--}_{-\frac12}
+\tilde d^{+-}_{-\frac32}\,\tilde d^{-+}_{-\frac12}\Big)\,|0\rangle_t,\nonumber
\end{eqnarray}
where the arrow in the second line refers to the series of spectral flows and coordinate maps and $\tilde\alpha_{A\dot A,m}$ and $\tilde d^{\alpha A}_m$ are the bosonic and fermionic operator modes natural to the covering space. We consider the bosonic part of $\partial_{w_0}|\chi\rangle$ in this subsection and analyze the term $\tilde\alpha_{+\dot+,-1}\tilde\alpha_{-\dot-,-1}$. The analysis for the other bosonic term $\tilde\alpha_{+\dot-,-1}\tilde\alpha_{-\dot+,-1}$ follows similarly.

Using the mode expansions in section \ref{sectiontwo} we have
\begin{equation}\label{alpha-1}
\tilde\alpha_{A\dot A,-1}=
\partial_tX_{A\dot A}(0)=
\frac{1}{2\pi}\oint_{t=0}\frac{dt}{t}\,\partial_tX_{A\dot A}(t).
\end{equation}
We map the complex plane to the cylinder via the map $z=e^w$. In the region close to the insertion of the deformation operator, $w_0=\tau_0+i\sigma_0$, we have
\begin{equation}\label{w0-1}
|z|=e^{\tau}=e^{\tau_0+\epsilon},\qquad\qquad\frac{\epsilon}{\tau_0}\ll1,
\end{equation}
where $\epsilon$ is a positive real number. On passing to the covering surface we have
\begin{eqnarray}\label{w0-2}
z=z_0+t^2 \Longrightarrow t\!\!\!\!\!\!&&
=\pm\,(z-z_0)^{\frac12}=\pm\Big(e^{\tau_0+\epsilon+i\sigma}-
e^{\tau_0+i\sigma_0}\Big)^{\frac12}\nonumber\\
&&=\pm e^{\frac{\tau_0}{2}+i\frac{\sigma_0}{2}}
\Big(e^{\epsilon+i(\sigma-\sigma_0)}-1\Big)^{\frac12}
\approx\pm e^{\frac{\tau_0}{2}+i\frac{\sigma_0}{2}}(w-w_0)^{\frac12}.
\end{eqnarray}
We then use this expression to rewrite the factor $1/t$ in the integrand of (\ref{alpha-1}) in terms of the cylinder coordinates $w$:
\begin{equation}\label{w0-3}
\frac{1}{t}=\frac{\pm e^{-\frac{w_0}{2}}}{(w-w_0)^{\frac12}}.
\end{equation}

Under the coordinate transformations which map the $t$ plane to the $w$ plane we have
\begin{equation}\label{w0-4}
\partial_tX(t)\,dt=\partial_wX(w)\,dw.
\end{equation}
We note that the bosons do not transform under the spectral flow transformations. Thus we only need to consider the conformal transformation properties of the boson fields in order to map them back to the cylinder. This is, however, not the case for fermions which carry $R$-charges and transform nontrivially under the spectral flows. We will discuss this in more detail in section \ref{fcont-i}.

For $\tau>\tau_0$ we have a single copy of the CFT on a doubly wound circle on the cylinder. The mode integrals on the cylinder in this region are then defined on this double circle. The positive and negative roots of (\ref{w0-2}) correspond to the two $2\pi$ turns around the doubly twisted circle. The contour integral (\ref{alpha-1}) in the cover then gives the sum over the two parts of the integral on the cylinder. Using (\ref{w0-3}) and (\ref{w0-4}) the contour integral (\ref{alpha-1}) reads
\begin{equation}\label{alpha-1-2}
\tilde\alpha_{A\dot A,-1}(0)=\oint_{t=0}\frac{dt}{2\pi t}\,\partial_{t}X_{A\dot A}(t)=
\int\frac{dw}{2\pi}e^{-\frac{w_0}{2}}\frac{\partial_wX_{A\dot A}(w)}{(w-w_0)^{\frac12}}.
\end{equation}

We now analyze the first bosonic term of $|\chi\rangle$ in (\ref{dsigma-2}) in the continuum limit. Under the series of spectral flows and coordinate transformations which map the covering surface to the cylinder, the NS vacuum $|0\rangle_t$ on the cover is mapped into the state $|\chi(w_0)\rangle$ on the cylinder. We then obtain
\begin{eqnarray}\label{alpha-1-3}
\frac{z_0}{2}\,\tilde\alpha_{+\dot+,-1}\tilde\alpha_{-\dot-,-1}|0\rangle_t\!\!\!\!\!\!\!&&=
\frac{z_0}{2}\,\oint_{t_1=0}\frac{dt_1}{2\pi t_1}\,\partial_{t_1}X_{+\dot+}(t_1)
\oint_{t_2=0}\frac{dt_2}{2\pi t_2}\,\partial_{t_2}X_{-\dot-}(t_2)\,|0\rangle_t\nonumber\\
&&\longrightarrow\frac{e^{w_0}}{2}\int\frac{dw_1}{2\pi}\,\frac{e^{-\frac{w_0}{2}}\partial_{w_1}X_{+\dot+}(w_1)}{(w_1-w_0)^{\frac12}}
\int\frac{dw_2}{2\pi}\,\frac{e^{-\frac{w_0}{2}}\partial_{w_2}X_{-\dot-}(w_2)}{(w_2-w_0)^{\frac12}}\,|\chi(w_0)\rangle\nonumber\\
&&=\int\frac{dw_1}{2\pi}\int\frac{dw_2}{2\pi}\,\partial_{w_1}X_{+\dot+}(w_1)\,\partial_{w_2}X_{-\dot-}(w_2)\,\mathcal{K}^B(w_1,w_2)\,|\chi(w_0)\rangle,
\end{eqnarray}
where the bosons are normal ordered and the arrow in the second line denotes the maps and spectral flows from the covering space to the cylinder. The kernel $\mathcal{K}^B$ is defined as
\begin{equation}\label{kernel-b-1}
\mathcal{K}^B(w_1,w_2)\equiv\frac12\,\frac{1}{(w_1-w_0)^{\frac12}\,(w_2-w_0)^{\frac12}}.
\end{equation}

The same results hold for the second bosonic term in (\ref{dsigma-2}), \emph{i.e.,} $\frac{z_0}{2}\,\tilde\alpha_{+\dot-,-1}\tilde\alpha_{-\dot+,-1}$. Thus we have
\begin{eqnarray}\label{alpha-1-4}
&&\frac{z_0}{2}\,\Big(-\tilde\alpha_{+\dot+,-1}\tilde\alpha_{-\dot-,-1}+\tilde\alpha_{+\dot-,-1}\tilde\alpha_{-\dot+,-1}\Big)\,|0\rangle_t\\
&&\longrightarrow\int\frac{dw_1}{2\pi}\int\frac{dw_2}{2\pi}\,\Big(-\partial_{w_1}X_{+\dot+}(w_1)\,\partial_{w_2}X_{-\dot-}(w_2)+
\partial_{w_1}X_{+\dot-}(w_1)\,\partial_{w_2}X_{-\dot+}(w_2)\Big)\times\nonumber\\
&&\qquad\qquad\qquad\qquad\qquad\qquad\qquad\qquad\qquad\qquad\qquad
\times\,\mathcal{K}^B(w_1,w_2)\,|\chi(w_0)\rangle.\nonumber
\end{eqnarray}

\subsection{Fermionic part of the squeezed state in the continuum limit}\label{fcont-i}
In this section we analyze the fermionic contribution to the squeezed state in the continuum limit. We consider the fermionic part of the state $|\chi\rangle$ (\ref{dsigma-2}) which is expressed in terms of the modes $\tilde d^{\alpha A}_r$ (\ref{psi-t-0-ii}) natural to the $t$ plane. We perform the calculations for the term $\tilde d^{++}_{-3/2}\tilde d^{--}_{-1/2}$. The analysis for the other term $\tilde d^{+-}_{-3/2}\tilde d^{-+}_{-1/2}$ will follow analogously.
The modes of fermions natural to the covering surface are given by
\begin{equation}\label{psi-t-0-ii}
\tilde d^{\alpha A}_{r}=\frac{1}{2\pi i}\oint_{t=0}dt\,t^{r-\frac12}\,\psi^{\alpha A}(t),
\quad r\in\mathbb{Z}+\frac12.
\end{equation}
Under the coordinate transformations which map the cover to the cylinder coordinates we have
\begin{equation}\label{dphidx}
\psi(t)\,dt=\Big(\frac{dt(w)}{dw}\Big)^{\frac12}\psi(w)\,dw.
\end{equation}
Using the expression we obtained in (\ref{w0-3}) we find that
\begin{equation}\label{dtdw}
\Big(\frac{dt(w)}{dw}\Big)^{\frac12}=\Big(\frac{dt}{dz}\,\frac{dz}{dw}\Big)^{\frac12}=
\frac1{\sqrt2}\,\frac{e^{\frac{w_0}{4}}}{(w-w_0)^{\frac14}}
\end{equation}
in the continuum limit.
For $\tilde d^{++}_{-3/2}$ and $\tilde d^{--}_{-1/2}$ we have
\begin{eqnarray}
&&\tilde d^{++}_{-\frac32}=\partial_t\psi^{++}(0)=
\oint_{t=0}\frac{dt}{2\pi i}\,\frac{\psi^{++}(t)}{t^2},\label{psi-t-3/2}\\
&&\tilde d^{--}_{-\frac12}=\psi^{--}(0)
=\oint_{t=0}\frac{dt}{2\pi i}\,\frac{\psi^{--}(t)}{t}.\label{psi-t-1/2}
\end{eqnarray}

Let us first consider $\tilde d^{++}_{-3/2}$ and use (\ref{dphidx}) and (\ref{w0-3}) to write (\ref{psi-t-3/2}) in terms of the $w$ coordinates. There is an important point to note here before we perform these calculations. In the process of computing the state $|\chi\rangle$ (\ref{dsigma-2}) we performed two spectral flow transformations: one on the cylinder with the spectral flow parameter $\alpha=-1$ and one on the cover with $\alpha=1$. Fermionic fields carry $SU(2)_R$ charges and acquire a phase under the spectral flow transformation. In order to express $\psi^{\alpha A}(t)$ in terms of $\psi^{\alpha A}(w)$ in the integrand of (\ref{psi-t-3/2}), one not only needs to take into account the conformal transformation properties of the fermion field, but also the spectral flow transformations on the cover and the cylinder. These two spectral flows have parameters $\alpha=-1$ on the cover and $\alpha=1$ on the cylinder. Let us first consider $\psi^{++}(t)$. We use (\ref{sf1bb}) and (\ref{sf2bb}) 
and find that $\psi^{++}(t)\to e^{\frac{w_0}{2}}t^{-\frac12}\psi^{++}(w)$. We then obtain
\begin{eqnarray}\label{psi-t-3/2-ii}
\tilde d^{++}_{-\frac32}\!\!\!\!\!\!&&=
\oint_{t=0}\frac{dt}{2\pi i}\,\frac{\psi^{++}(t)}{t^2}\nonumber\\
&&=\int\frac{dw}{2\pi i}\,\frac{\frac{1}{\sqrt2}\,e^{\frac{w_0}{4}}(w-w_0)^{-\frac14}
\psi^{++}(w)\,e^{\frac{w_0}{4}}\,(w-w_0)^{-\frac14}}{e^{w_0}(w-w_0)}\nonumber\\
&&=\frac{1}{\sqrt2}\,\int\frac{dw}{2\pi i}\,\frac{\psi^{++}(w)\,e^{-\frac{w_0}{2}}}
{(w-w_0)^{\frac32}}.
\end{eqnarray}
For $\psi^{--}(t)$ we have $\psi^{--}(t)\to e^{-\frac{w_0}{2}}t^{\frac12}\psi^{--}(w)$ and
\begin{eqnarray}\label{psi-t-1/2-ii}
\tilde d^{--}_{-\frac12}\!\!\!\!\!\!&&=
\oint_{t=0}\frac{dt}{2\pi i}\,\frac{\psi^{--}(t)}{t}\nonumber\\
&&=\int\frac{dw}{2\pi i}\,\frac{\frac{1}{\sqrt2}\,e^{\frac{w_0}{4}}(w-w_0)^{-\frac14}
\psi^{--}(w)\,e^{-\frac{w_0}{4}}\,(w-w_0)^{\frac14}}
{e^{\frac{w_0}{2}}(w-w_0)^{\frac12}}\nonumber\\
&&=\frac{1}{\sqrt2}\,\int\frac{dw}{2\pi i}\,\frac{\psi^{--}(w)\,e^{-\frac{w_0}{2}}}
{(w-w_0)^{\frac12}}.
\end{eqnarray}
We can now write the full expression for fermions. Under the spectral flows and coordinate transformations, the NS vacuum $|0\rangle_t$ on the cover is mapped to the state $|\chi(w_0)\rangle$ (\ref{dsigma-2}) on the cylinder. We then have
\begin{eqnarray}\label{dd-i}
\frac{z_0}{2}\,\tilde d^{++}_{-\frac32}\,\tilde d^{--}_{-\frac12}\,|0\rangle_t\!\!\!\!\!\!&&\longrightarrow
\frac{e^{w_0}}{2(\sqrt2)^2}\,\int\frac{dw_1}{2\pi i}\,\frac{\psi^{++}(w_1)\,e^{-\frac{w_0}{2}}}
{(w_1-w_0)^{\frac32}}\int\frac{dw_2}{2\pi i}\,\frac{\psi^{--}(w_2)\,e^{-\frac{w_0}{2}}}
{(w_2-w_0)^{\frac12}}\,|\chi(w_0)\rangle\nonumber\\
&&=\int \frac{dw_1}{2\pi i}\int \frac{dw_2}{2\pi i}\,\psi^{++}(w_1)\,\psi^{--}(w_2)\,{\mathcal{K}}^{\prime F}(w_1,w_2)\,|\chi(w_0)\rangle,
\end{eqnarray}
where the fermions are normal ordered and the fermion kernel ${\mathcal{K}}^{\prime F}$ is defined as
\begin{equation}\label{kernel-f-1}
{\mathcal{K}}^{\prime F}(w_1,w_2)\equiv
\frac{1}{4}\,\frac{1}{(w_1-w_0)^{\frac32}(w_2-w_0)^{\frac12}}.
\end{equation}

The boson kernel ${\mathcal{K}}^{B}$ (\ref{kernel-b-1}) and the fermion kernel in the above equation are related to each other as we now describe. The boson kernel $\mathcal{K}^B$ acts as $\partial X_{A\dot A}\,\mathcal{K}^B\,\partial X_{B\dot B}$ and the fermion kernel acts as $\psi^{\alpha A}\,{\mathcal{K}}^{\prime F}\,\psi^{\beta B}$. If we integrate the fermion kernel ${\mathcal{K}}^{\prime F}$ (\ref{kernel-f-1}) by parts over $w_1$, we get $\partial\psi^{\alpha A}\,\mathcal{K}^F\psi^{\beta B}$, where
\begin{equation}\label{kernelbf}
\mathcal{K}^F=\mathcal{K}^B=\frac{1}{2}\,\frac{1}{(w_1-w_0)^{\frac12}(w_2-w_0)^{\frac12}}.
\end{equation}
The kernels for $\partial X_{A\dot A}\,\mathcal{K}^B\,\partial X_{B\dot B}$ and $\partial\psi^{\alpha A}\,\mathcal{K}^F\,\psi^{\beta B}$ are thus the same.

The same results obtained above hold for the second fermionic term in (\ref{dsigma-2}): $\frac{z_0}{2}\,\tilde d^{+-}_{-3/2}\tilde d^{-+}_{-1/2}$. The total fermionic contribution reads
\begin{eqnarray}\label{dd-ii}
&&\frac{z_0}{2}\,\Big(-\tilde d^{++}_{-3/2}\tilde d^{--}_{-1/2}+\tilde d^{+-}_{-3/2}\tilde d^{-+}_{-1/2}\Big)\,|0\rangle_t\\
&&\longrightarrow\int\frac{dw_1}{2\pi}\int\frac{dw_2}{2\pi}\,\Big(\partial\psi^{++}(w_1)\,\psi^{--}(w_2)-\partial\psi^{+-}(w_1)\,\psi^{-+}(w_2)\Big)\times\nonumber\\
&&\qquad\qquad\qquad\qquad\qquad\qquad\qquad\qquad\qquad\qquad
\times\,\mathcal{K}^F(w_1,w_2)\,|\chi(w_0)\rangle.\nonumber
\end{eqnarray}

\subsection{The complete squeezed state in the continuum limit}\label{cont-tot}

It is now straightforward to combine the results of the bosonic and fermionic contributions (\ref{alpha-1-4}) and fermion (\ref{dd-ii}).  This gives
\begin{eqnarray}
&&\frac{z_0}{2}\,\Big(\tilde{L}_{-2}+\tilde{J}^3_{-2}\Big)\,|0\rangle_t\\
&&=\frac{z_0}{2}\,\Big(-\tilde\alpha_{+\dot+,-1}\,\tilde\alpha_{-\dot-,-1}+
\tilde\alpha_{+\dot-,-1}\,\tilde\alpha_{-\dot+,-1}
-\tilde d^{++}_{-\frac32}\,\tilde d^{--}_{-\frac12}
+\tilde d^{+-}_{-\frac32}\,\tilde d^{-+}_{-\frac12}\Big)\,|0\rangle_t\nonumber\\
&&\longrightarrow\Bigg[\int\frac{dw_1}{2\pi}\int\frac{dw_2}{2\pi}\,\Bigg(-\partial_{w_1}X_{+\dot+}(w_1)\,\partial_{w_2}X_{-\dot-}(w_2)+
\partial_{w_1}X_{+\dot-}(w_1)\,\partial_{w_2}X_{-\dot+}(w_2)+\nonumber\\
&&\qquad\qquad\quad
+\partial\psi^{++}(w_1)\,\psi^{--}(w_2)-\partial\psi^{+-}(w_1)\,\psi^{-+}(w_2)\Bigg)
\,\frac{\frac12}{(w_1-w_0)^{\frac12}(w_2-w_0)^{\frac12}}\Bigg]\,|\chi(w_0)\rangle.\nonumber
\end{eqnarray}
This yields a differential equation for the state $| \chi(w_0)\rangle$
\begin{eqnarray}
&& \partial_{w_0}| \chi(w_0)\rangle= \\
&&\Bigg[\int\frac{dw_1}{2\pi}\int\frac{dw_2}{2\pi}\,\Bigg(-\partial_{w_1}X_{+\dot+}(w_1)\,\partial_{w_2}X_{-\dot-}(w_2)+
\partial_{w_1}X_{+\dot-}(w_1)\,\partial_{w_2}X_{-\dot+}(w_2)+\nonumber\\
&&\qquad\qquad\quad
+\partial\psi^{++}(w_1)\,\psi^{--}(w_2)-\partial\psi^{+-}(w_1)\,\psi^{-+}(w_2)\Bigg)
\,\frac{\frac12}{(w_1-w_0)^{\frac12}(w_2-w_0)^{\frac12}}\Bigg]\,|\chi(w_0)\rangle.\nonumber
\end{eqnarray}
Note that the location of the twist insertion $w_0$ only appears in the kernel.  This allows us to integrate this expression, and find
\begin{eqnarray}\label{conttot}
&&|\chi(w_0)\rangle=\exp\Bigg[\int\frac{dw_1}{2\pi}\int\frac{dw_2}{2\pi}
\Bigg(\partial_{w_1}X_{+\dot+}(w_1)\,\partial_{w_2}X_{-\dot-}(w_2)-\partial_{w_1}X_{+\dot-}(w_1)\,\partial_{w_2}X_{-\dot+}(w_2)+\nonumber\\
&&\qquad\quad-\partial_{w_1}\psi^{++}(w_1)\psi^{--}(w_2)+\partial_{w_1}\psi^{+-}(w_1)\psi^{-+}(w_2)\Bigg)
\ln{\big(\sqrt{w_1-w_0}+\sqrt{w_2-w_0}\big)}\Bigg]|0^-_R\rangle,\nonumber\\
\end{eqnarray}
up to an overall normalization.  This normalization is fixed by the earlier convention that $\sigma_2^+(w_0)|0^-_R\rangle^{(1)} \otimes | 0^-_R\rangle^{(2)} =| 0^{-}_R\rangle +\cdots$.  A second way to arrive at this state is by directly analyzing the sums in the exponentials of the squeezed state (\ref{pfive}), and replacing these sums with integrals in the continuum limit.  We do this in appendix \ref{altkernels}.
\section{Discussion}

If we could understand the D1D5 CFT away from the orbifold point, then we would be able to unravel many mysteries about the black hole. In particular, we could ask how the black hole forms from the infalling matter, or how it reacts when additional matter falls in; this is expected to correspond to a thermalization process in the dual CFT. The CFT at the orbifold point is essentially free and so does not thermalize. It is hoped that with a study of the deformations away from the orbifold point we would be able to get a qualitative picture of the thermalization process.

Since thermalization is likely to involve many orders of deformation from the orbifold point, it is useful to have a good understanding of the effect of the deformation operator. In the present paper we have seen that while the coefficients $\gamma^B, \gamma^F$ look complicated, their {\it derivative} with respect to $z_0$ is in fact simple: in the covering space it has the schematic form  $\t\alpha_{-1}\t\alpha_{-1}+ \t d_{-\frac{3}{2}} \t d_{-\h}$.

We have also investigated the squeezed state produced by the twist deformation in the continuum limit. In a position space representation, the exponential in the squeezed state is given through kernels $K^B(\sigma_1, \sigma_2), K^F(\sigma_1, \sigma_2)$. Looking at these kernels gives a useful physical picture of the effect of the twist; one can see the correlations produced in the fields around the location of the twist insertion.

In future work, we hope to return to the results here to extract a general qualitative picture of deforming away from the orbifold point as well as to apply the stress-tensor method to higher twist states.

\section*{Acknowledgements}

We wish to thank Steve Avery, Zachary Carson, Borun Chowdhury, Shaun Hampton, and David Turton for many helpful comments. The work of SDM is supported in part by DOE grant DE-SC0011726.  The research of AWP was supported by the Natural Sciences and Engineering Research Council (NSERC) of Canada. IGZ is supported by the DOE grant DE-SC0009987 and by the NSF grant PHY-1053842.

\appendix
\renewcommand\theequation{\thesection.\arabic{equation}}
\setcounter{equation}{0}
\section{Notation and the CFT algebra} \label{appa}

In the D1D5 CFT at the orbifold point, we have four real left moving fermions $\psi^1, \psi^2, \psi^3, \psi^4$.  We group these into complex doublets $\psi^{\alpha A}$
\be
\begin{pmatrix} \psi^{++} \\ \psi^{-+}\end{pmatrix}=\sqi\begin{pmatrix}\psi_1+i\psi_2\\ \psi_3+i\psi_4\end{pmatrix}
\label{aaone}
\ee
\be
\begin{pmatrix} \psi^{+-}\\  \psi^{--}\end{pmatrix}=\sqi\begin{pmatrix} \psi_3-i\psi_4\\  -(\psi_1-i\psi_2) \end{pmatrix}
\label{aatwo}
\ee
The first index, which we label $\alpha=(+,-)$, denotes the transformation properties under $SU(2)_L$, which is a subgroup of rotations on $S^3$.  The second index, which we label $A=(+,-)$, is an index of the subgroup $SU(2)_1$ of rotations in $T^4$. These four complex fermions have a reality condition because they are constructed from only four real fermions:
\be
 (\psi^\dagger)_{\alpha A}=-\epsilon_{\alpha\beta}\epsilon_{AB} \psi^{\beta B}.
\ee
The two-point function for these complex fermions are
\be
\langle\psi^{\alpha A}(z)\psi^\dagger_{\beta B}(w)\rangle=\delta^\alpha_\beta\delta^A_B{\frac{1}{z-w}}, ~~~
\langle\psi^{\alpha A}(z)\psi^{\beta B}(w)\rangle=-\epsilon^{\alpha\beta}\epsilon^{AB}{\frac{1}{z-w}}
\ee
where the $\epsilon$ symbol is defined as
\be
\epsilon_{12}=1, ~~~\epsilon^{12}=-1, ~~~
\psi_A=\epsilon_{AB}\psi^B, ~~~
\psi^A=\epsilon^{AB}\psi_B.
\ee
In addition to the fermions, there are four real left moving bosons $X_1, X_2, X_3, X_4$ which can be grouped into a matrix
\be
X_{A\dot A}= \sqi X_i \sigma^i=\sqi \begin{pmatrix} X_3+iX_4 & X_1-iX_2 \\ X_1+iX_2&-X_3+iX_4 \end{pmatrix}
\label{aathree}
\ee
where $\sigma^i$ for $i=1,2,3$ are the usual Pauli matrices, and $\sigma^4=i{\mathbb I}$.  The complex conjugate of the above fields are given by
\be
(X^*)^{A\dot A}=\sqi \begin{pmatrix} X_3-iX_4& X_1+iX_2 \\
X_1-iX_2&-X_3-iX_4 \end{pmatrix}
\ee
leading again to a reality condition
\be
(X_{A\dot A})^*=X^{A\dot A}
\ee
which is expected because we constructed four complex bosons from four real bosons, so there must be a condition of this sort to reduce the number of degrees of freedom.  From these definitions, the 2-point functions are
\be
\langle \p X_{A\dot A}(z) (\p X^\dagger)^{B\dot B}(w)\rangle=-{\frac{1}{(z-w)^2}}\delta^B_A\delta^{\dot B}_{\dot A}, ~~~
\langle\p X_{A\dot A}(z) \p X_{B\dot B}(w)\rangle={\frac{1}{(z-w)^2}}\epsilon_{AB}\epsilon_{\dot A\dot B}
\ee

The symmetry currents
\be
J^a=-{\frac{1}{4}}(\psi^\dagger)_{\alpha A} (\sigma^{Ta})^\alpha{}_\beta \psi^{\beta A}
\label{Jop}
\ee
\be
G^\alpha_{\dot A}= \psi^{\alpha A} \p X_{A\dot A}, ~~~(G^\dagger)_{\alpha}^{\dot A}=(\psi^\dagger)_{\alpha A} \p (X^\dagger)^{A\dot A}
\label{Gop}
\ee
\be
T=-{\h} (\p X^\dagger)^{A\dot A}\p X_{A\dot A}-{\h} (\psi^\dagger)_{\alpha A} \p \psi^{\alpha A}
\label{Top}
\ee
\be
(G^\dagger)_{\alpha}^{\dot A}=-\epsilon_{\alpha\beta} \epsilon^{\dot A\dot B}G^\beta_{\dot B}, ~~~~G^{\alpha}_{\dot A}=-\epsilon^{\alpha\beta} \epsilon_{\dot A\dot B}(G^\dagger)_\beta^{\dot B}
\ee
obey the following OPE algebra
\be
J^a(z) J^b(w)\sim \delta^{ab} {\frac{\h}{(z-w)^2}}+i\epsilon^{abc} {\frac{J^c}{z-w}}
\ee
\be
J^a(z) G^\alpha_{\dot A} (z')\sim {\frac{1}{(z-z')}}\h (\sigma^{aT})^\alpha{}_\beta G^\beta_{\dot A}
\ee
\be
G^\alpha_{\dot A}(z) (G^\dagger)^{\dot B}_\beta(z')\sim -{\frac{2}{(z-z')^3}}\delta^\alpha_\beta \delta^{\dot B}_{\dot A}- \delta^{\dot B}_{\dot A}  (\sigma^{Ta})^\alpha{}_\beta [{\frac{2J^a}{(z-z')^2}}+{\frac{\p J^a}{(z-z')}}]
-{\frac{1}{(z-w)}}\delta^\alpha_\beta \delta^{\dot B}_{\dot A}T
\ee
\be
T(z)T(z')\sim {\frac{3}{(z-z')^4}}+{\frac{2T}{(z-z')^2}}+{\frac{\p T}{(z-z')}}
\ee
\be
T(z) J^a(z')\sim {\frac{J^a}{(z-z')^2}}+{\frac{\p J^a}{(z-z')}}
\ee
\be
T(z) G^\alpha_{\dot A}\sim \frac{{\frac{3}{2}}G^\alpha_{\dot A}}{(z-z')^2}  + \frac{\p G^\alpha_{\dot A}}{(z-z')}.
\ee

Working on the fundamental $\psi$ fields, note that
\be
J^a(z) \psi^{\gamma C}(w)\sim {\h} \frac{1}{ z-w} (\sigma^{aT})^\gamma{}_\beta \psi^{\beta C}
\ee

The above OPEs lead to the following algebra for modes of the symmetry currents
\begin{eqnarray}
\com{J^a_m}{J^b_n} &=& \frac{m}{2}\delta^{ab}\delta_{m+n,0} + i{\epsilon^{ab}}_c J^c_{m+n}
            \\
\com{J^a_m}{G^\alpha_{\dot{A},n}} &=& \frac{1}{2}{(\sigma^{aT})^\alpha}_\beta G^\beta_{\dot{A},m+n}
             \\
\ac{G^\alpha_{\dot{A},m}}{G^\beta_{\dot{B},n}} &=& \hspace*{-4pt}\epsilon_{\dot{A}\dot{B}}\bigg[
   (m^2 - \frac{1}{4})\epsilon^{\alpha\beta}\delta_{m+n,0}
  + (m-n){(\sigma^{aT})^\alpha}_\gamma\epsilon^{\gamma\beta}J^a_{m+n}
  + \epsilon^{\alpha\beta} L_{m+n}\bigg]\\
\com{L_m}{L_n} &=& \frac{m(m^2-\frac{1}{4})}{2}\delta_{m+n,0} + (m-n)L_{m+n}\\
\com{L_m}{J^a_n} &=& -n J^a_{m+n}\\
\com{L_m}{G^\alpha_{\dot{A},n}} &=& \left(\frac{m}{2}-n\right)G^\alpha_{\dot{A},m+n}
\end{eqnarray}.


\section{Spectral flow}\label{appb}
Consider a state $|\psi_{h,j}\rangle$ with dimension $h$ and charge $j$. Under spectral flow by a parameter $\alpha$, this state changes to a different state with dimension $h'$ and charge $j'$ \cite{spectral}:
\be
|\psi_{h,j}\rangle~\r~|\psi'_{h',j'}\rangle,
\ee
where
\be
h'=h+\alpha j +{\frac{\alpha^2 c}{ 24}},
\label{atwo}
\ee
\be
j'=j+\frac{\alpha c}{12}.
\label{aone}
\ee
Consider a  primary operator ${\cal O}_j$ carrying charge $j$ on the cylinder with coordinate $w$. Under spectral flow this operator changes as
\be
{\cal O}_j ~\r~ {\cal O}'_j~=~{\cal O}_j e^{-\alpha j w}+\dots.
\label{asix}
\ee
The operator $J$ is  neutral, but is modified by spectral flow due to the current anomaly. The operator $T$ is not a primary, and also has the conformal anomaly. Our goal is to see how these operators change under spectral flow.

\subsection{The changes in  \texorpdfstring{$J_0$}{Lg} and \texorpdfstring{$L_0$}{Lg}}

The current operator $J_0$ is neutral, but changes by an additive constant due to the current anomaly:
\be
J_0~\r ~ J'_0~=~ J_0+a_1.
\ee
We can find $a_1$ by using the change in charge under spectral flow. We have
\be
J_0|\psi_{h,j}\rangle=j |\psi_{h,j}\rangle.
\ee
If we spectral flow {\it both} the state and the charge operator, we will get the same eigenvalue
\be
J'_0|\psi'_{h',j'}\rangle=(J_0+a_1)|\psi'_{h',j'}\rangle=j |\psi'_{h',j'}\rangle.
\ee
From (\ref{aone}) we have
\be
J_0|\psi'_{h',j'}\rangle=(j+\frac{\alpha c}{12}) |\psi'_{h',j'}\rangle.
\label{athree}
\ee
Thus we get
$a_1=-{\frac{\alpha c}{12}}$ and the operator $J_0$ transforms as
\be
J_0~\r ~ J'_0~=~ J_0-\frac{\alpha c}{12}.
\ee

Let us now consider the energy operator $L_0$, for which we can follow a similar procedure.
The change in $L_0$ has the form
\be
L_0~\r ~ L'_0~=~ L_0+a_2 J_0+a_3,
\ee
where $a_2, a_3$ are constants. We have
\be
L_0|\psi_{h,j}\rangle=h |\psi_{h,j}\rangle.
\ee
If we spectral flow {\it both} the state and the energy operator, we will get the same eigenvalue
\be
L'_0|\psi'_{h',j'}\rangle=(L_0+a_2 J_0+a_3)|\psi'_{h',j'}\rangle=h |\psi'_{h',j'}\rangle.
\ee
Using  (\ref{athree}), this gives
\be
(L_0+a_2 (j+{\frac{\alpha c}{12}})+a_3)|\psi'_{h',j'}\rangle=h |\psi'_{h',j'}\rangle.
\label{afour}
\ee
From (\ref{atwo}) we have
\be
L_0|\psi'_{h',j'}\rangle=(h+\alpha j + \frac{\alpha^2 c}{24}) |\psi'_{h',j'}\rangle.
\label{afive}
\ee
Substituting  (\ref{afive}) in (\ref{afour}) we find
\be
(h+\alpha j + \frac{\alpha^2 c}{24})+a_2 (j+\frac{\alpha c}{12})+a_3=h.
\ee
Thus we get $a_2=-\alpha$, $a_3=\frac{\alpha ^2 c}{24}$,  and the operator $L_0$ transforms as
\be
L_0~\r ~ L'_0~=~ L_0-\alpha J_0+\frac{\alpha^2 c}{24}.
\ee

\subsection{The changes in \texorpdfstring{$J(w)$}{Lg} and \texorpdfstring{$T(w)$}{Lg} on the cylinder}

From the shifts of $J_0, L_0$ under spectral flow, we can find the shifts of the operators $J(w), T(w)$ on the cylinder. We have
\be
J_0=\frac{1}{2\pi} \int_{\sigma=0}^{2\pi} dw J(w),\qquad L_0=\frac{1}{2\pi}\int_{\sigma=0}^{2\pi} dw T(w).
\ee
Thus
\be
J(w)~\r~J'(w)=J(w)-\frac{\alpha c}{12},
\ee
\be
T(w)~\r~T'(w)=T(w)-\alpha J(w)+\frac{\alpha^2 c}{24}.
\ee

\subsection{The changes in \texorpdfstring{$J(z)$}{Lg} and \texorpdfstring{$T(z)$}{Lg} on the plane}

We will also need to find the effect of spectral flow on operators on the complex plane. Consider the plane $z$ defined through
$z=e^w$. The analog of (\ref{asix}) is
\be
{\cal O}_j ~\r~ {\cal O}'_j~=~{\cal O}_j z^{-\alpha j }+\dots.
\ee

Consider the operator
 $J(z)$, and spectral flow by parameter $\alpha$ around the origin $z=0$. To find the change in $J(z)$, we proceed in three steps:

\b

($i$) We map $J(z)$ to the cylinder, getting
\be
J(w)=\left ( \frac{\p z}{\p w}\right )J(z) = z J(z).
\ee

\b

($ii$) We perform the spectral flow by parameter $\alpha$ on the cylinder, getting
\be
J(w) ~\r ~ J(w) -\frac{\alpha c}{12}.
\ee

\b

($iii$) We then map back to the plane, getting
\bea
J'(z)&=&\left ( \frac{\p w}{\p z}\right )\left  [J(w) -\frac{\alpha c}{12}\right ]= \frac{1}{z}\left [ z J(z) -\frac{\alpha c}{12}\right ]\nn
&=& J(z)-\frac{\alpha c}{12 z}.
\eea

\b

We can perform the same steps to find the change in $T(z)$;

\b

($i$) We map $T(z)$ to the cylinder, getting
\be
T(w)=\left ( \frac{\p z}{\p w}\right )^2 T(z) +\frac{c}{12} \{ z,w \}=z^2 T(z) -\frac{1}{4}.
\ee

\b

($ii$) We perform the spectral flow by parameter $\alpha$, getting
\be
T(w)~\r~T(w)-\alpha J(w)+\frac{\alpha^2 c}{24}.
\ee

\b

($iii$) We map back to the plane, getting
\bea
T'(z)&=&\left ( \frac{\p w}{\p z}\right )^2 \left  [T(w)-\alpha J(w)+\frac{\alpha^2 c}{24}\right ] + \frac{c}{12} \{ w,z \} \nn
&=& \frac{1}{z^2} \left [ z^2 T(z) -\frac{1}{4}-\alpha z J(z)+\frac{\alpha^2 c}{24}\right ]+\frac{1}{4 z^2}\nn
&=& T(z) -\frac{\alpha J(z)}{z}+\frac{\alpha^2 c}{24 z^2}.
\eea

\b

\subsection{Spectral flows used in our computations}

Let us now see how we use the above relations in our computations in section \ref{iii-i}.

\b

(a) On the cylinder we perform a spectral flow transformation with parameter $\alpha=1$. Using $c=6$ we find equation (\ref{aten}):
\be
T'(w)=T(w)-J(w)+\frac{1}{4}.
\ee

(b) On the $t$-plane we spectral flow by $\alpha=-1$. This then gives equation (\ref{sone}):
\be
T'(t)=T(t)+\frac{J(t)}{t}+\frac{1}{4 t^2}
\ee


\section{Directly analyzing the squeezed state} \label{altkernels}

\subsection{bosonic contribution}\label{bcont-ii}
The results obtained in section \ref{sectionfour} for the continuum limit can be obtained by performing computations directly on the cylinder using $|\chi(w_0)\rangle$ (\ref{pfive}):
\begin{eqnarray}\label{chichi-2}
&&|\chi(w_0)\rangle=\sigma_2^+(w_0)\,|0^-_R\rangle^{(1)}\otimes|0^-_R\rangle^{(2)}=\\
&&e^{\sum_{m\ge1,n\ge1}\gamma^B_{mn}
\left(-\alpha_{+\dot+,-m}\alpha_{-\dot-,-n}+\alpha_{-\dot+,-m}\alpha_{+\dot-,-n} \right)}\,
e^{\sum_{m\ge0,n\ge1}\gamma^F_{mn}
\left(d^{++}_{-m}d^{--}_{-n}-d^{+-}_{-m}d^{-+}_{-n} \right)}|0^-_R\rangle.\nonumber
\end{eqnarray}
We focus on the bosonic part in this section. Let us consider the bosonic term
\begin{equation}\label{chichi-3}
e^{-\sum_{m^\prime\ge0,n^\prime\ge0}\gamma^B_{2m^\prime+1,2n^\prime+1}
\alpha_{+\dot+,-(2m^\prime+1)}\alpha_{-\dot-,-(2n^\prime+1)}},
\end{equation}
where we have set $m=2m'+1$ and $n=2n'+1$ because the coefficients $\gamma^B_{mn}$ and $\gamma^F_{mn}$ are nonzero only for odd values of $m$ and $n$ (see section \ref{sectiontwo}).  The bosonic modes $\alpha_{A\dot A,m}$ are defined on the doubly wound circle in the region $\tau>\tau_0$ above the insertion point of the twist operator.  These modes are given by
\begin{equation}\label{mode-cyl-bos-af-2}
\alpha_{A\dot A,m}=\frac{1}{2\pi}\int_{\sigma}^{\sigma+4\pi}dw\,
\partial_wX_{A\dot A}(w)\,e^{\frac m2w}.
\end{equation}
We then obtain
\begin{equation}\label{delX-cyl-af-1}
\partial_wX_{A\dot A}=-\frac{i}{2}\sum_{m}\alpha_{A\dot A,m}e^{-\frac{m}{2}w}.
\end{equation}
In what follows, it will be more convenient to write the modes in terms of $X$ rather than $\partial X$.  We write, with a caveat,
\begin{equation}\label{delX-cyl-af-2}
X_{A\dot A}(w)=\frac{i}{2}\sum_{m\neq0}\frac{2}{m}\,
\alpha_{A\dot A,m}e^{-\frac{m}{2}w}.
\end{equation}
This is not the full $X$ operator.  Rather, it is the holomorphic part of $X$ with both the center of mass mode $x_{A \dot{A}}$ and the momentum mode $\alpha_{A \dot{A},0}$ removed.  It would be sufficient to simply remove the center of mass mode, given that this operator acts on the zero momentum Ramond vacuum.  We suppress this subtlety here, writing $X$ to simplify notation in this section.  At the end, we will integrate by parts to put the derivative back on the $X$ operator, which automatically removes the center of mass mode $x$, and because the operator is acting on the Ramond vacuum, the momentum modes evaluate to $0$.  This will reinstate the $\partial X$ as being the full operator.  Thus, in our final answer, we will be using conventional notation.

This subtlety being noted, we write the odd modes $m=2m^\prime+1$ in (\ref{mode-cyl-bos-af-2}) we find
\begin{eqnarray}\label{mode-cyl-bos-af-3}
\alpha_{A\dot A,m}\!\!\!\!\!\!&&=
\frac{1}{2\pi}\int_{\sigma}^{\sigma+4\pi}dw\,
\big(\frac{-m}{2}\big)X_{A\dot A}(w)\,e^{\frac m2w}\nonumber\\
&&=\frac{1}{2\pi}\int_{\sigma}^{\sigma+4\pi}dw\,
\big(m^\prime+\frac12\big)X_{A\dot A}(w)\,e^{-(m^\prime+\frac12)w}.
\end{eqnarray}
We use the above expression and rewrite the exponent of (\ref{chichi-3})
\begin{eqnarray}\label{chichi-4}
&&-\sum_{m^\prime\ge0,n^\prime\ge0}\gamma^B_{2m^\prime+1,2n^\prime+1}
\alpha_{+\dot+,-(2m^\prime+1)}\alpha_{-\dot-,-(2n^\prime+1)}=\nonumber\\
&&-\int_{\sigma_1}^{\sigma_1+4\pi}\frac{dw_1}{2\pi}\int_{\sigma_2}^{\sigma_2+4\pi}\frac{dw_2}{2\pi}\,
X_{+\dot+}(w_1)X_{-\dot-}(w_2)\times\nonumber\\
&&\qquad\qquad\times\sum_{m^\prime\ge0,n^\prime\ge0}\gamma^B_{2m^\prime+1,2n^\prime+1}
\big(m^\prime+\frac12\big)\big(n^\prime+\frac12\big)\,
e^{-(m^\prime+\frac12)w_1}e^{-(n^\prime+\frac12)w_2},
\end{eqnarray}
where
\begin{equation}\label{gammab-2}
\gamma^B_{2m^\prime+1,2n^\prime+1}=
\frac{2\,z_0^{(m^\prime+n^\prime+1)}}
{\pi\,(2m^\prime+1)(2n^\prime+1)(1+m^\prime+n^\prime)}\;
\frac{\Gamma[\frac32+m^\prime]\,\Gamma[\frac32+n^\prime]}
{\Gamma[1+m^\prime]\,\Gamma[1+n^\prime]}.
\end{equation}
In the continuum limit, $\frac{z_0}{z_1}\to1,\frac{z_0}{z_2}\to1$, $m^\prime\to\infty,n^\prime\to\infty$, and the summation is replaced with integrals over $dm^\prime$ and $dn^\prime$. In this limit we obtain
\begin{eqnarray}\label{gammab-3}
\gamma^B_{2m^\prime+1,2n^\prime+1}\!\!\!\!\!\!&&\approx
\frac{1}{2\pi}\,\frac{z_0^{(m^\prime+n^\prime+1)}}{m^\prime+n^\prime}\;
\frac{\Gamma[\frac12+m^\prime]}{\Gamma[1+m^\prime]}\,
\frac{\Gamma[\frac12+n^\prime]}{\Gamma[1+n^\prime]}
\end{eqnarray}
Using properties of the Gamma functions
\begin{equation}\label{gamma-1}
\lim_{m\to\infty}\frac{\Gamma(m+\beta)}{\Gamma(m)\,m^\beta}=1
\quad\beta\in\mathbb{R},
\end{equation}
allows us to approximate the coefficients $\gamma^B_{mn}$ (\ref{gammab-3}).  We then replace the remaining sums over $m'$ and $n'$ as integrals in (\ref{chichi-4}) giving
\begin{eqnarray}\label{chichi-5}
&&-\sum_{m^\prime\ge0,n^\prime\ge0}\gamma^B_{2m^\prime+1,2n^\prime+1}
\alpha_{+\dot+,-(2m^\prime+1)}\alpha_{-\dot-,-(2n^\prime+1)}\approx\nonumber\\
&&-\int_{\sigma_1}^{\sigma_1+4\pi}\frac{dw_1}{2\pi}\int_{\sigma_2}^{\sigma_2+4\pi}\frac{dw_2}{2\pi}\,
X_{+\dot+}(w_1)X_{-\dot-}(w_2)\times\nonumber\\
&&\qquad\times\int_{0}^{\infty}dm\int_{0}^{\infty}dn\,
\frac{1}{2\pi}\,\frac{z_0^{m^\prime+n^\prime+1}}{m^\prime+n^\prime}\;
\frac{m^\prime\,n^\prime}{m^{\prime{\frac12}}\,{n^\prime}^{\frac12}}
e^{-(m^\prime+\frac12)w_1}e^{-(n^\prime+\frac12)w_2}
\end{eqnarray}
recalling that we have taken $m'$ and $n'$ to be large.  We switch to coordinates
\begin{equation}\label{mpm-1}
m_+\equiv m^\prime+n^\prime,\qquad\qquad m_-\equiv m^\prime-n^\prime.
\end{equation}
Thus, our approximation becomes
\begin{eqnarray}\label{chichi-6}
&&-\sum_{m^\prime\ge0,n^\prime\ge0}\gamma^B_{2m^\prime+1,2n^\prime+1}
\alpha_{+\dot+,-(2m^\prime+1)}\alpha_{-\dot-,-(2n^\prime+1)}\approx\nonumber\\
&&-\int_{\sigma_1}^{\sigma_1+4\pi}\frac{dw_1}{2\pi}\int_{\sigma_2}^{\sigma_2+4\pi}\frac{dw_2}{2\pi}\;
X_{+\dot+}(w_1)X_{-\dot-}(w_2)\times\nonumber\\
&&\qquad\times\frac{1}{2}\int_{0}^{\infty}dm_+\,\frac{z_0^{m_++1}}{4\pi\,m_+}
e^{-\frac12(m_++1)w_+}\int_{-m_+}^{m_+}dm_-\,
\big(m_+^2-m_-^2\big)^{\frac12}\,e^{-\frac12m_-w_-},
\end{eqnarray}
where $w_+$ and $w_-$ are defined as
\begin{equation}
w_+\equiv w_1+w_2, \qquad\qquad w_-\equiv w_1-w_2.
\end{equation}
We first evaluate the integral over $dm_-$, which gives a Bessel function as the solution.  We find
\be
\int_{-m_+}^{m_+}dm_-\,\big(m_+^2-m_-^2\big)^{\frac12}\,e^{-\frac12m_-w_-}=
2\pi\,\frac{m_+}{w_-}\,I_1\big(\frac{m_+w_-}{2}\big).
\ee
Inserting this result back in (\ref{chichi-6}) we obtain the following integral over $m_+$
\begin{eqnarray}\label{chichi-7}
&&\frac{1}{2}\int_{0}^{\infty}dm_+\,\frac{z_0^{m_++1}}{4\pi\,m_+}
e^{-\frac12(m_++1)w_+}\int_{-m_+}^{m_+}dm_-\,
\big(m_+^2-m_-^2\big)^{\frac12}\,e^{-\frac12m_-w_-}\nonumber\\
&&\qquad\qquad=\frac{e^{-\frac12(w_+-2w_0)}}{4\,w_-}\int_{0}^{\infty}dm_+\,
e^{-\frac12\,(w_+-2w_0)\,m_+}\,I_1\big(\frac{w_-m_+}{2}\big).
\end{eqnarray}
We use the table of integrals for the modified Bessel functions and consider the expression
\begin{equation}\label{modbessel-2}
\int_0^\infty\,dx\,e^{-\alpha x}\,I_\nu(\beta x)=\frac{\beta^{-\nu}\,
\Big(\alpha-\sqrt{\alpha^2-\beta^2}\Big)^\nu}{\sqrt{\alpha^2-\beta^2}},\qquad\qquad
\mathrm{Re}\,\nu>-1,\quad\mathrm{Re}\,\alpha>|\mathrm{Re}\,\beta|.
\end{equation}
We identify $\alpha=(w_+-2w_0)/2$, $\beta=w_-/2$.  The first condition for the integral above is clearly satisfied because $\nu=1$.  For the second, we recall that we are working in a region just above the insertion of the twist operator, and so
\begin{eqnarray}
&&w_0=\tau_0+i\sigma_0,\quad\qquad w_1=\tau_0+\epsilon_1+i\sigma_1,
\qquad w_2=\tau_0+\epsilon_2+i\sigma_2,\label{omegas-1}\\
&&w_+=2\tau_0+(\epsilon_1+\epsilon_2)+i(\sigma_1+\sigma_2),\qquad
w_-=(\epsilon_1-\epsilon_2)+i(\sigma_1-\sigma_2)\label{omegas-1-1}
\end{eqnarray}
with $\epsilon_1>0, \epsilon_2>0$.  Hence,
\begin{equation}\label{omegas-2}
\alpha=\frac{(\epsilon_1+\epsilon_2)}{2}+i\frac{(\sigma_1+\sigma_2-2\sigma_0)}{2},
\qquad \beta=\frac{(\epsilon_1-\epsilon_2)}{2}+i\frac{(\sigma_1-\sigma_2)}{2}.
\end{equation}
and the second condition $\mathrm{Re}\,\alpha>|\mathrm{Re}\,\beta|$ is met as well.  The integral over $dm_+$ then reads
\begin{eqnarray}
\int_{0}^{\infty}dm_+\,
e^{-\frac12\,(w_+-2w_0)m_+}\,I_1\big(\frac{w_-m_+}{2}\big)&&\kern -1.5em =
\frac{2}{w_-}\bigg(\frac{1}{2}\frac{\big(\sqrt{w_1-w_0}-\sqrt{w_2-w_0}\big)^2}
{\sqrt{w_1-w_0}\,\sqrt{w_2-w_0}}\bigg)\nonumber  \\
&&\kern -8em =\frac{w_{-}}{\big(\sqrt{w_1-w_0}+\sqrt{w_2-w_0}\big)^2}
\frac{1}{\sqrt{w_1-w_0}\,\sqrt{w_2-w_0}}\label{intdm+2}
\end{eqnarray}
after some algebraic simplification.  Thus
\begin{eqnarray}\label{intdm+3}
&& \frac{e^{-\frac12(w_+-2w_0)}}{4\,w_-}\int_{0}^{\infty}dm_+\,z_0^{m_++1}\,
I_1\big(\frac{m_+w_-}{2}\big)\,e^{-\frac12(m_++1)\,w_+}\\
&&\qquad =\frac14\,
e^{-\frac12\big[(w_1-w_0)+(w_2-w_0)\big]}
\frac{1}{\big(\sqrt{w_1-w_0}+\sqrt{w_2-w_0}\big)^2}
\frac{1}{\sqrt{w_1-w_0}\,\sqrt{w_2-w_0}} \nonumber \\
&&\qquad \approx \frac{1}{\big(\sqrt{w_1-w_0}+\sqrt{w_2-w_0}\big)^2}
\frac{1}{\sqrt{w_1-w_0}\,\sqrt{w_2-w_0}}.\nonumber
\end{eqnarray}
where in the last expression, we use that $z_0/z_1\to 1$ and $z_0/z_2\to 1$ and so the exponential becomes 1 in the continuum limit.

We use this to compute the exponent (\ref{chichi-5}). Therefore, in the continuum limit we obtain
\begin{eqnarray}\label{chichi-8}
&&-\sum_{m^\prime\ge0,n^\prime\ge0}\gamma^B_{2m^\prime+1,2n^\prime+1}
\alpha_{+\dot+,-(2m^\prime+1)}\alpha_{-\dot-,-(2n^\prime+1)}\approx\nonumber\\
&&-\int_{\sigma_1}^{\sigma_1+4\pi}\frac{dw_1}{2\pi}\int_{\sigma_2}^{\sigma_2+4\pi}\frac{dw_2}{2\pi}\,
X_{+\dot+}(w_1)X_{-\dot-}(w_2)\,\hat{\mathcal{K}}^B(w_1,w_2),
\end{eqnarray}
where the kernel is of the form
\begin{equation}\label{kernelb-dir}
\hat{\mathcal{K}}^B(w_1,w_2)\equiv\frac14\,
\frac{1}{(w_1-w_0)^{\frac12}\,(w_2-w_0)^{\frac12}}
\frac{1}{\big[(w_1-w_0)^{\frac12}+(w_2-w_0)^{\frac12}\big]^2}.
\end{equation}
In section \ref{cont-tot-cyl} we will see how the kernel $\hat{\mathcal{K}}^B$ obtained above through direct computations on the cylinder is related to the kernel ${\mathcal{K}}^B$ (\ref{kernel-b-1}) obtained in the main text.  Essentially, this comes down to integrating by parts to put the derivatives back on $X(w_1)$ and $X(w_2)$.  As mentioned earlier in this section, this will be important to bring us back to more conventional notation.  The calculation for the other bosonic contribution is identical, and so we find the total bosonic contribution to be
\begin{eqnarray}\label{chichi-9-i}
\sum_{m^\prime\ge0,n^\prime\ge0}\gamma^B_{2m^\prime+1,2n^\prime+1}\,
\Big(-\alpha_{+\dot+,-(2m^\prime+1)}\,\alpha_{-\dot-,-(2n^\prime+1)}+ \alpha_{+\dot-,-(2m^\prime+1)}\,\alpha_{-\dot+,-(2n^\prime+1)}\Big)\approx\nonumber\\
\int_{\sigma_1}^{\sigma_1+4\pi}\frac{dw_1}{2\pi}\int_{\sigma_2}^{\sigma_2+4\pi}\frac{dw_2}{2\pi}\,
\Big(-X_{+\dot+}(w_1)\,X_{-\dot-}(w_2)+X_{+\dot-}(w_1)\,X_{-\dot+}(w_2)\Big)\,\hat{\mathcal{K}}^B(w_1,w_2).
\end{eqnarray}

\subsection{fermionic contribution}\label{fcont-ii}
In this subsection we consider the fermion part of the state $|\chi(w_0)\rangle$ (\ref{pfive}) on the cylinder in the continuum limit. The fermion part is of the form
\begin{equation}\label{chichi-9}
e^{\sum_{m^\prime\ge0,n^\prime\ge0}\gamma^F_{2m^\prime+1,2n^\prime+1}
\big(d^{++}_{-(2m^\prime+1)}d^{--}_{-(2n^\prime+1)}-
d^{+-}_{-(2m^\prime+1)}d^{-+}_{-(2n^\prime+1)}\big)},
\end{equation}
where the modes of fermions $d^{\alpha A}_{m}$ are defined on the doubly twisted circle in the region $\tau>\tau_0$ and are given by
\begin{equation}\label{mode-cyl-fer-af-2}
d^{\alpha A}_{m}=\frac{1}{2\pi i}\int_{\sigma}^{\sigma+4\pi}dw\,
\psi^{\alpha A}(w)\,e^{\frac m2w}.
\end{equation}
We then find
\begin{equation}\label{fer-cyl-af-1}
\psi^{\alpha A}(w)=\frac12\sum_md^{\alpha A}_m\,e^{-\frac m2w}.
\end{equation}
For the odd modes $m=2m^\prime+1$ in (\ref{mode-cyl-fer-af-2}) we find
\begin{equation}\label{mode-cyl-fer-af-3}
d^{\alpha A}_{m}=
\int_{\sigma}^{\sigma+4\pi}\frac{dw}{2\pi i}\,
\psi^{\alpha A}(w)\,e^{-(m^\prime+\frac 12)\,w}.
\end{equation}
We write the exponent of (\ref{chichi-9}) using the above relation. We perform the computations for the term $d^{++}_{-(2m^\prime+1)}d^{--}_{-(2n^\prime+1)}$ and note the calculations for the other term $d^{+-}_{-(2m^\prime+1)}d^{-+}_{-(2n^\prime+1)}$ follow analogously. The exponent is given by
\begin{eqnarray}\label{chichi-10}
&&\sum_{m^\prime\ge0,n^\prime\ge0}\gamma^F_{2m^\prime+1,2n^\prime+1}
d^{++}_{-(2m^\prime+1)}d^{--}_{-(2n^\prime+1)}=\\
&&\int_{\sigma_1}^{\sigma_1+4\pi}\frac{dw_1}{2\pi i}
\int_{\sigma_2}^{\sigma_2+4\pi}\frac{dw_2}{2\pi i}\,\psi^{++}(w_1)\psi^{--}(w_2)\!\!\!\!\!\!
\sum_{m^\prime\ge0,n^\prime\ge0}\gamma^F_{2m^\prime+1,2n^\prime+1}
e^{-(m^\prime+\frac12)w_1}e^{-(n^\prime+\frac12)w_2},\nonumber
\end{eqnarray}
where
\begin{equation}\label{gammaf-2}
\gamma^F_{2m^\prime+1,2n^\prime+1}=
-\frac{\,z_0^{(m^\prime+n^\prime+1)}}
{\pi\,(2n^\prime+1)(1+m^\prime+n^\prime)}\;
\frac{\Gamma[\frac32+m^\prime]\,\Gamma[\frac32+n^\prime]}
{\Gamma[1+m^\prime]\,\Gamma[1+n^\prime]}.
\end{equation}

In the continuum limit, $\{\frac{z_0}{z_1},\frac{z_0}{z_2}\}\to1$, and $\{m^\prime,n^\prime\}\to\infty$.  Again, applying the property of gamma matrices (\ref{gamma-1}) we find
\begin{equation}\label{gammab-4}
\gamma^F_{2m^\prime+1,2n^\prime+1}\approx
-\frac{1}{2\pi}\,\frac{z_0^{(m^\prime+n^\prime+1)}}{m^\prime+n^\prime}\;
\frac{m^{\prime{\frac12}}}{{n^\prime}^{\frac12}}.
\end{equation}
After replacing the sums with integrals, the exponent (\ref{chichi-10}) then reads
\begin{eqnarray}\label{chichi-11}
\sum_{m^\prime\ge0,n^\prime\ge0}\gamma^F_{2m^\prime+1,2n^\prime+1}
d^{++}_{-(2m^\prime+1)}d^{--}_{-(2n^\prime+1)}
\int_{\sigma_1}^{\sigma_1+4\pi}\frac{dw_1}{2\pi i}\int_{\sigma_2}^{\sigma_2+4\pi}\frac{dw_2}{2\pi i}\,
\psi_{++}(w_1)\psi_{--}(w_2)\times\nonumber\\
\times\int_{0}^{\infty}dm^\prime\int_{0}^{\infty}dn^\prime\,
\frac{(-1)}{2\pi}\,\frac{z_0^{m^\prime+n^\prime+1}}{m^\prime+n^\prime}\;
\frac{{m^\prime}^{\frac12}}{n^{\prime{\frac12}}}
e^{-(m^\prime+\frac12)w_1}e^{-(n^\prime+\frac12)w_2}.\qquad
\end{eqnarray}
We want to evaluate the integrals over $m^\prime$ and $n^\prime$ in the second line. Let us rewrite the second line of the above expression as
\begin{eqnarray}\label{expint-1}
\mathcal{I}\!\!\!\!\!\!&&\equiv\int_{0}^{\infty}dm^\prime\int_{0}^{\infty}dn^\prime\,
\frac{(-1)}{2\pi}\,\frac{z_0^{m^\prime+n^\prime+1}}{m^\prime+n^\prime}\;
\frac{{m^\prime}^{\frac12}}{n^{\prime{\frac12}}}
e^{-(m^\prime+\frac12)w_1}e^{-(n^\prime+\frac12)w_2}\nonumber\\
&&=-\frac{1}{2\pi}\,e^{-\frac{(w_1+w_2)}{2}}e^{w_0}
\int_{0}^{\infty}dm^\prime\,m^{\prime\frac12}\,e^{-(w_1-w_0)m^\prime}
\int_{0}^{\infty}dn^\prime\,\frac{1}{n^{\prime\frac12}(n^\prime+m^\prime)}\,
e^{-(w_2-w_0)n^\prime}.
\end{eqnarray}

We first evaluate the integral over $n^\prime$. Using table of integrals for exponentials we consider the following expression
\begin{equation}\label{expint-2}
\int_0^\infty dx\,\frac{x^{\nu-1}\,e^{-\mu x}}{x+\beta}=
\beta^{\nu-1}\,e^{\beta\mu}\,\Gamma(\nu)\,\Gamma(1-\nu,\beta\mu),\qquad
|\mathrm{arg}\,\beta|<\pi, \mathrm{Re}\,\mu>0, \mathrm{Re}\,\nu>0,
\end{equation}
where $\Gamma(\eta,x)$ is the upper incomplete Gamma function.
In our case, $\nu=1/2$, $\mu=w_2-w_0$, $\beta=m^\prime$, and so the conditions in (\ref{expint-2}) are satisfied. The integral over $dn^\prime$ in (\ref{expint-1}) then reads
\begin{eqnarray}\label{expint-3}
\int_{0}^{\infty}dn^\prime\,\frac{1}{n^{\prime\frac12}(n^\prime+m^\prime)}\,
e^{-(w_2-w_0)n^\prime}\!\!\!\!\!\!&&=\frac{\pi^\frac12}{m^{\prime\frac12}}\,e^{(w_2-w_0)m^\prime}\,
\Gamma\Big(\frac12,(w_2-w_0)m^\prime\Big).
\end{eqnarray}
Putting this expression back in (\ref{expint-1}) we obtain
\begin{eqnarray}\label{expint-4}
\mathcal{I}\!\!\!\!\!\!&&=-\frac{\pi^\frac12}{2\pi}\,e^{-\frac{(w_1+w_2)}{2}}e^{w_0}
\int_{0}^{\infty}dm^\prime\,e^{-(w_1-w_2)m^\prime}\,
\Gamma\Big(\frac12,(w_2-w_0)m^\prime\Big).
\end{eqnarray}
We define a new parameter $m^{\prime\prime}\equiv(w_2-w_0)\,m^{\prime}$. The above integral then reads
\begin{equation}\label{expint-5}
\mathcal{I}=-\frac{1}{2\pi^\frac12}\,e^{-\frac12\big((w_1-w_0)+(w_2-w_0)\big)}
\frac{1}{(w_2-w_0)}\int_{0}^{\infty}dm^{\prime\prime}\,
e^{-\frac{w_1-w_2}{w_2-w_0}\,m^{\prime\prime}}\,\Gamma\big(\frac12,m^{\prime\prime}\big).
\end{equation}
We next use the table of integrals for the incomplete Gamma functions to evaluate the integral over $m^{\prime\prime}$. We have
\begin{equation}\label{expint-6}
\int_0^\infty dx\,e^{-\alpha x}\,\Gamma(\xi,x)=\frac{1}{\alpha}\,\Gamma(\xi)
\left[1-\frac{1}{(1+\alpha)^\beta}\right],\qquad\qquad\xi>0.
\end{equation}
Comparing this expression to (\ref{expint-5}) we find that $\alpha=(w_1-w_2)/(w_2-w_0)$, and $\xi=1/2>0$. We then obtain
\begin{eqnarray}\label{expint-7}
\mathcal{I}\!\!\!\!\!\!&&=-\frac12\,e^{-\frac12\big[(w_1-w_0)+(w_2-w_0)\big]}\,
\frac{1}{\sqrt{w_1-w_0}(\sqrt{w_1-w_0}+\sqrt{w_2-w_0)})}\\
&& \approx -\frac12
\frac{1}{\sqrt{w_1-w_0}(\sqrt{w_1-w_0}+\sqrt{w_2-w_0)})}
\end{eqnarray}
where we have used the continuum limit $z_0/z_1\to1$ and $z_0/z_2\to1$ to remove the preceding exponentials.  Inserting this into the exponent (\ref{chichi-10}), we find
\begin{eqnarray}\label{chichi-14}
\sum_{m^\prime\ge0,n^\prime\ge0}\gamma^F_{2m^\prime+1,2n^\prime+1}
d^{++}_{-(2m^\prime+1)}d^{--}_{-(2n^\prime+1)}\approx\qquad\qquad\qquad\nonumber\\
\int_{\sigma_1}^{\sigma_1+4\pi}\frac{dw_1}{2\pi i}\int_{\sigma_2}^{\sigma_2+4\pi}\frac{dw_2}{2\pi i}\,\psi^{++}(w_1)\psi^{--}(w_2)\,\hat{\mathcal{K}}^F(w_1,w_2),
\end{eqnarray}
where the kernel is given by
\begin{equation}\label{kernel-f-dir}
\hat{\mathcal{K}}^F(w_1,w_2)\equiv-\frac12\,\frac{1}{(w_1-w_0)^{\frac12}}\;\frac{1}{[(w_1-w_0)^{\frac12}+(w_2-w_0)^{\frac12}]}.
\end{equation}
In section \ref{cont-tot-cyl} we will compare the fermion kernel $\hat{\mathcal{K}}^F$ obtained in the above equation to the fermion kernel ${\mathcal{K}}^F$ (\ref{kernelbf}) computed in the main text. As mentioned before, the kernel for the other fermionic term $d^{+-}_{-(2m^\prime+1)}d^{-+}_{-(2n^\prime+1)}$ is the same as (\ref{kernel-f-dir}). Therefore, the fermionic contribution to the squeezed state in the continuum limit is given by
\begin{eqnarray}\label{chichi-15}
\sum_{m^\prime\ge0,n^\prime\ge0}\gamma^F_{2m^\prime+1,2n^\prime+1}\,
\Big(d^{++}_{-(2m^\prime+1)}d^{--}_{-(2n^\prime+1)}-d^{+-}_{-(2m^\prime+1)}d^{-+}_{-(2n^\prime+1)}\Big)\approx\qquad\qquad\qquad\nonumber\\
\int_{\sigma_1}^{\sigma_1+4\pi}\frac{dw_1}{2\pi i}\int_{\sigma_2}^{\sigma_2+4\pi}\frac{dw_2}{2\pi i}\,
\Big(\psi^{++}(w_1)\psi^{--}(w_2)-\psi^{+-}(w_1)\psi^{-+}(w_2)\Big)\,\hat{\mathcal{K}}^F(w_1,w_2).
\end{eqnarray}


\subsection{Squeezed state on the cylinder}\label{cont-tot-cyl}

In subsetions \ref{bcont-ii} and \ref{fcont-ii} we considered the squeezed state on the cylinder (\ref{chichi-2}) and investigated the bosonic and fermionic contributions in the continuum limit. Combining the bosonic (\ref{chichi-9-i}) and fermionic (\ref{chichi-15}) parts, we obtain the full expression
\begin{eqnarray}\label{conttotcyl-1}
&&\kern -2em |\chi(w_0)\rangle=\sigma_2^+(w_0)\,|0^-_R\rangle^{(1)}\otimes|0^-_R\rangle^{(2)}=\\
&&\kern -2em e^{\sum_{m\ge1,n\ge1}\gamma^B_{mn}
\left(-\alpha_{+\dot+,-m}\alpha_{-\dot-,-n}+\alpha_{-\dot+,-m}\alpha_{+\dot-,-n} \right)}\,
e^{\sum_{m\ge0,n\ge1}\gamma^F_{mn}
\left(d^{++}_{-m}d^{--}_{-n}-d^{+-}_{-m}d^{-+}_{-n} \right)}\,|0^-_R\rangle\approx\nonumber\\
&& \kern -2em \quad e^{\int_{\sigma_1}^{\sigma_1+4\pi}\frac{dw_1}{2\pi}\int_{\sigma_2}^{\sigma_2+4\pi}\frac{dw_2}{2\pi}\,
\Big(-X_{+\dot+}(w_1)\,X_{-\dot-}(w_2)+X_{+\dot-}(w_1)\,X_{-\dot+}(w_2)\Big)\,
\frac{\frac14}{\sqrt{w_1-w_0}\,\sqrt{w_2-w_0}}\,\frac{1}{(\sqrt{w_1-w_0}+\sqrt{w_2-w_0})^2}}\nonumber\\
&&\times\,e^{\int_{\sigma_1}^{\sigma_1+4\pi}\frac{dw_1}{2\pi i}\int_{\sigma_2}^{\sigma_2+4\pi}\frac{dw_2}{2\pi i}\,
\Big(\psi^{++}(w_1)\psi^{--}(w_2)-\psi^{+-}(w_1)\psi^{-+}(w_2)\Big)\,
\frac{\frac{(-1)}{2}}{\sqrt{w_1-w_0}}\;\frac{1}{(\sqrt{w_1-w_0}+\sqrt{w_2-w_0})}}\,|0^-_R\rangle.\nonumber
\end{eqnarray}
In order to compare this expression to (\ref{conttot}), we perform integration by parts over $w_1$ and $w_2$ for the bosonic terms and over $w_1$ for the fermionic terms. We then find
\begin{eqnarray}\label{conttotcyl-1-ii}
&&\kern -2em |\chi(w_0)\rangle=\\
&&\kern -2em e^{\int_{\sigma_1}^{\sigma_1+4\pi}\frac{dw_1}{2\pi}\int_{\sigma_2}^{\sigma_2+4\pi}\frac{dw_2}{2\pi}\,
\Big(-\partial_{w_1}X_{+\dot+}(w_1)\,\partial_{w_2}X_{-\dot-}(w_2)+\partial_{w_1}X_{+\dot-}(w_1)\,\partial_{w_2}X_{-\dot+}(w_2)\Big)\,
(-\ln{(\sqrt{w_1-w_0}+\sqrt{w_2-w_0})})}\nonumber\\
&&\kern -2em \quad\times\,e^{\int_{\sigma_1}^{\sigma_1+4\pi}\frac{dw_1}{2\pi i}\int_{\sigma_2}^{\sigma_2+4\pi}\frac{dw_2}{2\pi i}\,
\Big(\partial_{w_1}\psi^{++}(w_1)\psi^{--}(w_2)-\partial_{w_1}\psi^{+-}(w_1)\psi^{-+}(w_2)\Big)\,(-\ln{(\sqrt{w_1-w_0}+\sqrt{w_2-w_0})})}\,|0^-_R\rangle.\nonumber
\end{eqnarray}
We note that the boson and fermion kernels are the same after performing integration by parts
\begin{equation}\label{kernel-cyl-ibp}
\hat{\mathcal{K}}^{\prime B}(w_1,w_2)=\hat{\mathcal{K}}^{\prime F}(w_1,w_2)\equiv
\ln{\Big(\sqrt{w_1-w_0}+\sqrt{w_2-w_0}\Big)}.
\end{equation}
We then have
\begin{eqnarray}\label{conttotcyl-2}
&&|\chi(w_0)\rangle=\exp\Bigg[\int_{\sigma_1}^{\sigma_1+4\pi}\frac{dw_1}{2\pi}\int_{\sigma_2}^{\sigma_2+4\pi}\frac{dw_2}{2\pi}\,\times\\
&&\times\,\Bigg(\partial_{w_1}X_{+\dot+}(w_1)\,\partial_{w_2}X_{-\dot-}(w_2)-\partial_{w_1}X_{+\dot-}(w_1)\,\partial_{w_2}X_{-\dot+}(w_2)+\nonumber\\
&&\quad-\partial_{w_1}\psi^{++}(w_1)\psi^{--}(w_2)+\partial_{w_1}\psi^{+-}(w_1)\psi^{-+}(w_2)\Bigg)
\ln{\Big(\sqrt{w_1-w_0}+\sqrt{w_2-w_0}\Big)}\Bigg]\,|0^-_R\rangle\nonumber
\end{eqnarray}
thus yielding the same result as the main text.


\end{document}